\documentclass[preprint,3p,authoryear,a4paper]{elsarticle}  
 
\usepackage{graphicx}
\usepackage[table]{xcolor}
\usepackage{dsfont}
\usepackage{amsmath}
\usepackage{amssymb}
\usepackage{amsfonts}
\usepackage{amsbsy}
\usepackage{amsthm}
\usepackage{array}
\usepackage{booktabs} 
\usepackage{verbatim} 
\usepackage[english]{babel}
\usepackage{float}
\usepackage{epsfig}
\usepackage{multirow}
\usepackage{epstopdf}
\usepackage{natbib}
\AtBeginDocument{}
\usepackage{pdfpages}
\usepackage{titlesec}
\usepackage{booktabs, array}
\usepackage{lipsum, tabularx}
\usepackage{siunitx,lipsum}
\usepackage{nameref}
\usepackage{color}
\usepackage{lscape}
\usepackage{tabularx}
\usepackage{subcaption, ulem}
\usepackage{soul}
\usepackage[toc,page]{appendix}

\renewcommand\appendix{\par
	\setcounter{section}{0}%
	\setcounter{subsection}{0}%
	\setcounter{table}{0}
	\setcounter{table}{0}
	\setcounter{figure}{0}
	\gdef\thetable{\Alph{table}}
	\gdef\thefigure{\Alph{figure}}
	\gdef\thesection{\Alph{section}}
	\setcounter{section}{0}}

\newcolumntype{L}[1]{>{\raggedright\let\newline\\\arraybackslash\hspace{0pt}}m{#1}}
\newcolumntype{C}[1]{>{\centering\let\newline\\\arraybackslash\hspace{0pt}}m{#1}}
\newcolumntype{R}[1]{>{\raggedleft\let\newline\\\arraybackslash\hspace{0pt}}m{#1}}

\renewcommand{\thefigure}{\arabic{section}.\arabic{figure}}

\renewcommand{\thetable}{\arabic{section}.\arabic{table}}

\newcommand{\rev}[1]{\textcolor{black}{#1}}

\numberwithin{equation}{section}

\newtheorem{remark}{Remark}[section]
\numberwithin{equation}{section}	

\newenvironment{arcitem}{
	\begin{list}{--}{
			\topsep=1pt 
			\itemsep=1pt 
			\parsep=0pt 
			\leftmargin=19pt 
	}}
	{\end{list}}

\newcounter{arclist}

\newcounter{arcenum}

\makeatletter
\def\ps@pprintTitle{%
	\let\@oddhead\@empty
	\let\@evenhead\@empty
	\def\@oddfoot{}%
	\let\@evenfoot\@oddfoot}
\makeatother
\begin{document}



	\begin{frontmatter}
		
		\title{On unbalanced data and common shock models in stochastic loss reserving}

		\author[UMelb]{Benjamin Avanzi}
		
		\author[UNSW]{Greg Taylor}
		
		\author[TF]{Phuong Anh Vu\corref{cor}}
		
		\author[UNSW]{Bernard Wong}
		
		\cortext[cor]{Correspondence to: Phuong Anh Vu, Taylor Fry, Level 22, 45 Clarence St, Sydney NSW 2000, Australia. E-mail: anh.vu@taylorfry.com.au}
		
		\address[UMelb]{Centre for Actuarial Studies, Department of Economics, University of Melbourne VIC 3010, Australia}
        \address[UNSW]{School of Risk and Actuarial Studies, Business School UNSW Sydney NSW 2052, Australia}
		\address[TF]{Taylor Fry, Level 22, 45 Clarence St, Sydney NSW 2000, Australia}
		
		\begin{abstract}
			 
		 Introducing common shocks is a popular dependence modelling approach, with some recent applications in loss reserving. The main advantage of this approach is the ability to capture structural dependence coming from known relationships. In addition, it helps with the parsimonious construction of correlation matrices of large dimensions. However, complications arise in the presence of ``unbalanced data", that is, when (expected) magnitude of observations over a single triangle, or between triangles, can vary substantially. Specifically, if a single common shock is applied to all of these cells, it can contribute insignificantly to the larger values and/or swamp the smaller ones, unless careful adjustments are made. This problem is further complicated in applications involving negative claim amounts. In this paper, we address this problem in the loss reserving context using a common shock Tweedie {approach for unbalanced data}. We show that the solution not only provides a much better balance of the common shock proportions relative to the unbalanced data, but it is also parsimonious.  Finally, the common shock Tweedie model also provides distributional tractability.

		\end{abstract}


	\end{frontmatter}

	\section{Introduction}\label{Sec:Intro}
	 Outstanding claims reserves are typically some of the most critical components  in the financial statement  of a non-life insurer \citep*{AbBoCo15,HeTh16,SaGi14}.  When estimating  reserves, the insurer often has to provide the central estimate as well as a risk margin to accommodate for the stochastic nature of outstanding claims. The estimation of reserving variability is also required by many regulators \citep*{GiJaMa12}. For example, the Australian Prudential Regulation Authority (APRA) requires insurers to provide a risk margin calculated as the larger of a half of one standard deviation, and the difference between 75th percentile and the expected value of the total outstanding claims distribution. The 99.5th percentile of the distribution of total outstanding claims is also an input in the calculation of risk based capital for solvency purposes in many regulatory frameworks, for example, Solvency II in Europe and APRA's Prudential Standards in Australia. This was one of the motivations for the development of stochastic reserving methodologies since the early 1980s. For general references on reserving, one can refer to \citet*{Tay00} and \citet*{WuMe08}. {A recent strand of the literature focuses on the modelling of individual claims (see for example, \citealp{AvWoYa16,PiAnDe03,Wut18,ZhZhJi09}. However, the focus of this paper is on the modelling of traditional aggregate data in the form of loss triangles.}
	
	Non-life insurers  typically operate in multiple lines or segments, and are required by regulators to estimate loss reserves and risk capital on an aggregate level. Different business lines  within an insurer's operation often lack a comonotonic dependence structure.  This allows the insurer to enjoy diversification benefits in  the calculation of loss reserves and risk capital for their consolidated  operation \citep*{AvTaWo16}. It is hence essential to develop an accurate approach to  the modelling of outstanding losses  while allowing for dependencies \citep*{CoGeAb16,ShBaMe12}. This  not only allows  the insurer to accurately assess their performance, but  also to hold an appropriate amount of reserves and capital to  optimise its internal use while satisfying regulatory requirements  \citep*{Ajn94,AvTaWo18}.
	
	Various multivariate approaches have been developed for stochastic loss reserving which take into account the dependency across business lines or segments. Some well-known non-parametric approaches include multivariate chain ladder frameworks in \citet*{Bra04,Sch06,MeWu07,Zha10} and the multivariate additive loss reserving framework in \citet*{HeScZo06,MeWu09}. These approaches are non-parametric  and do not utilise any distributional assumptions. They also focus on specific cell-wise dependence (i.e. the dependence between cells that are in the same position)  across loss triangles. Alternatively, parametric approaches utilising distributional assumptions can be used, see for example, \citet*{ShFr11,ZhDu13,Dej12,AbBoCo15,Shi14}.
	
	In this paper, we focus on the common shock approach to dependence modelling. Common shock approaches use common random factors to capture drivers of dependence across related variables. As a result, these drivers can be identified, as well as monitored if needed. The transparent dependence structures in common shock  models can then be interpreted more easily. This is indeed one of the four desirable properties of multivariate distributions considered in \citet*[Chapter 4]{Joe97} which include:
		\begin{arcitem}
			\item  interpretability,
			\item  closure under the taking of marginals, meaning that the multivariate marginals belong to the same family (this is important if, in modelling, we need to first choose appropriate univariate marginals, then bivariate and sequentially to higher-order marginals),
			\item  flexible and wide range of dependence,
			\item  density and cumulative distribution function in closed-form (if not, they are computationally feasible to work with).
		\end{arcitem} 
		Furthermore, the construction of correlation matrices can be facilitated. Correlation matrices are tools extensively used by practitioners to specify dependence in the aggregation of outstanding claims liabilities or risk-based capital. Explicit dependence structures captured using common shock approaches allow correlation matrices to be specified in a more disciplined and parsimonious manner \citep*[see e.g.,][]{AvTaWo18}. 
	 
	{Common shock approaches have been used to good effect}. They are typically used to capture structural dependence, that is, ``structural co-movements that are due to known relationships which can be accounted for in a modelling exercise" \citep*{IAA04}. \citet*{Dej06} introduced three different models to capture dependence across development periods, accident periods and calendar periods respectively. Calendar period dependence is captured using common shock variables in the multivariate log-normal model of \citet*{ShBaMe12}. A common shock Tweedie framework was developed in \citet*{AvTaVuWo16} to capture cell-wise dependence across business lines. {It is worth noting that these models are static models which assume a single development pattern for all accident years through the use of fixed effects. A recent use of common shock approach in evolutionary reserving models which allow claims development pattern to evolve can be found in \citet{AvTaVuWo19}. There are also various applications of common shock models outside of the reserving literature, including mortality modelling \citep{AlLaSh13,AlLaSh16}, capital modelling \citep{FuLa10} and claim counts modelling \citep{Mey07}}.
	
	{Despite the benefits mentioned above, complications can arise in the application of common shock approaches to loss triangle data. This is due to the ``unbalanced" feature of data where expected magnitude of observations within a loss triangle as well as across triangles vary substantially. This feature represents the typical claim experience where the level of {claim activity} reaches a peak in early years then dies out as the development lag increases. The ``unbalanced-ness" can also be observed in loss data that consists of multiple business lines. In particular, the speed of claims development can vary across business lines where some lines are longer tailed than others. As a result, the magnitude of claim observations in the same accident year and development year can vary across loss triangles. Because of this feature, we say that loss reserving data is an example of ``unbalanced data". If a single common shock is applied to these observations that are of different magnitudes, it can contribute insignificantly to the {larger} ones and/or swamp the smaller ones, unless careful adjustments are made. It is the aim of this paper to study and address this problem.}
	
	{While this paper aims to examine the challenges for common shock models and propose a solution to address these challenges, a focus of the solution is placed on the Tweedie family of distributions. This is motivated by its popularity. This family is a major subclass of the exponential dispersion family (EDF) consisting of symmetric and non-symmetric, light-tailed and heavy-tailed distributions \citep{AlLaSh152,Jor97}. Various members of it have been frequently used in the loss reserving literature, see for example, \citet{AlWu09,BoDa11,EnVe02,PeShWu09,ReVe98,Tay09,Tay15,Wut03,ZhDuGu12}. \citet{AvTaWo16} developed a common shock Tweedie framework for reserving to allow for dependence across business line while utilising the flexibility of this family of distribution. The solution proposed in this paper will be illustrated using this framework.}\par

	Another feature that is occasionally observed in loss triangles are negative claim amounts. These are due to various reasons, for example, salvage recoveries, or payment from third parties. Many common used distributions such as gamma distributions and log-normal distributions are unable to handle this feature due to their lack of support for negative values. A remarkably small area of literature has been devoted for the treatment of negative payments a single business line. The existing methods include a three-parameter-log-normal model in \citet{Dea06} and a mixture model in \citet{Kun06}. In the development of the new approach for unbalanced data, we will also consider a treatment for negative claims.\par

	The organisation of this paper is as follows: Section \ref{Sec:Challenges} investigates the issue of unbalanced data for common shock models. {A common shock Tweedie approach to unbalanced data is introduced in Section \ref{Sec:Model}}. Simulation illustrations are provided in Section \ref{Sec:sim}, including an illustration using a portfolio of triangles with different tail lengths, and a comparison of the performances of the original common shock Tweedie approach and the modified Tweedie approach with treatment for unbalanced data. An illustration using real data is provided in Section \ref{Sec:realdata} and Section \ref{Sec:Conclusion} concludes the paper.
	
\section{{Unbalanced feature of reserving data and its challenges to common shock models}} \label{Sec:Challenges}
{In this section, we examine the unbalanced feature of loss reserving data in detail. The general common shock framework developed in \citet{AvTaWo18} is then described. Challenges that arise in applying common shock models to reserving data due to its unbalanced feature are then discussed.}
	
\subsection{{Unbalanced feature of reserving data}}\label{Sec:unbalancednature}
{As described in Section \ref{Sec:Intro}, loss reserving data typically exhibits unbalanced nature. We consider for illustration a real data set from a Canadian insurer collected from 2003 to 2012 (denoted by years 1-10). This data set is used for illustration in \citet{CoGeAb16} and is provided in Tables \ref{Tab:real1} and \ref{Tab:real2} in Appendix \ref{Sec:real}. The two lines of business used for illustration are Bodily Injury line and Accident Benefits (excluding Disability Income).}\par 

\begin{figure}[H]
	\begin{subfigure}{.45\textwidth}
		\includegraphics[scale=0.5]{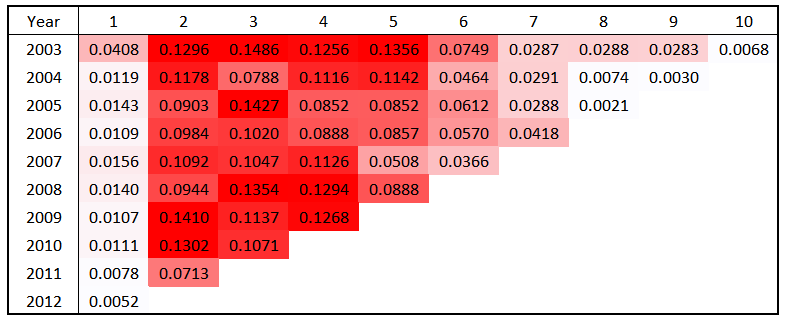}\\
		\includegraphics[scale=0.5]{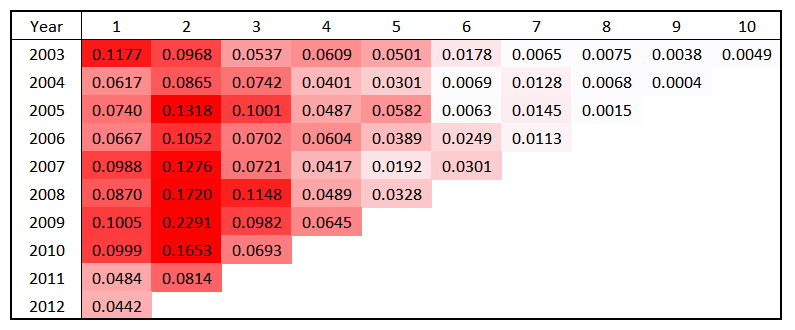}
		\subcaption{Heat maps of Bodily Injury line (top) and Accident Benefit line (bottom)}
	\end{subfigure}
	\hspace{1.4cm} 
	\begin{subfigure}{.45\textwidth}
		\centering
		\includegraphics[scale=0.42]{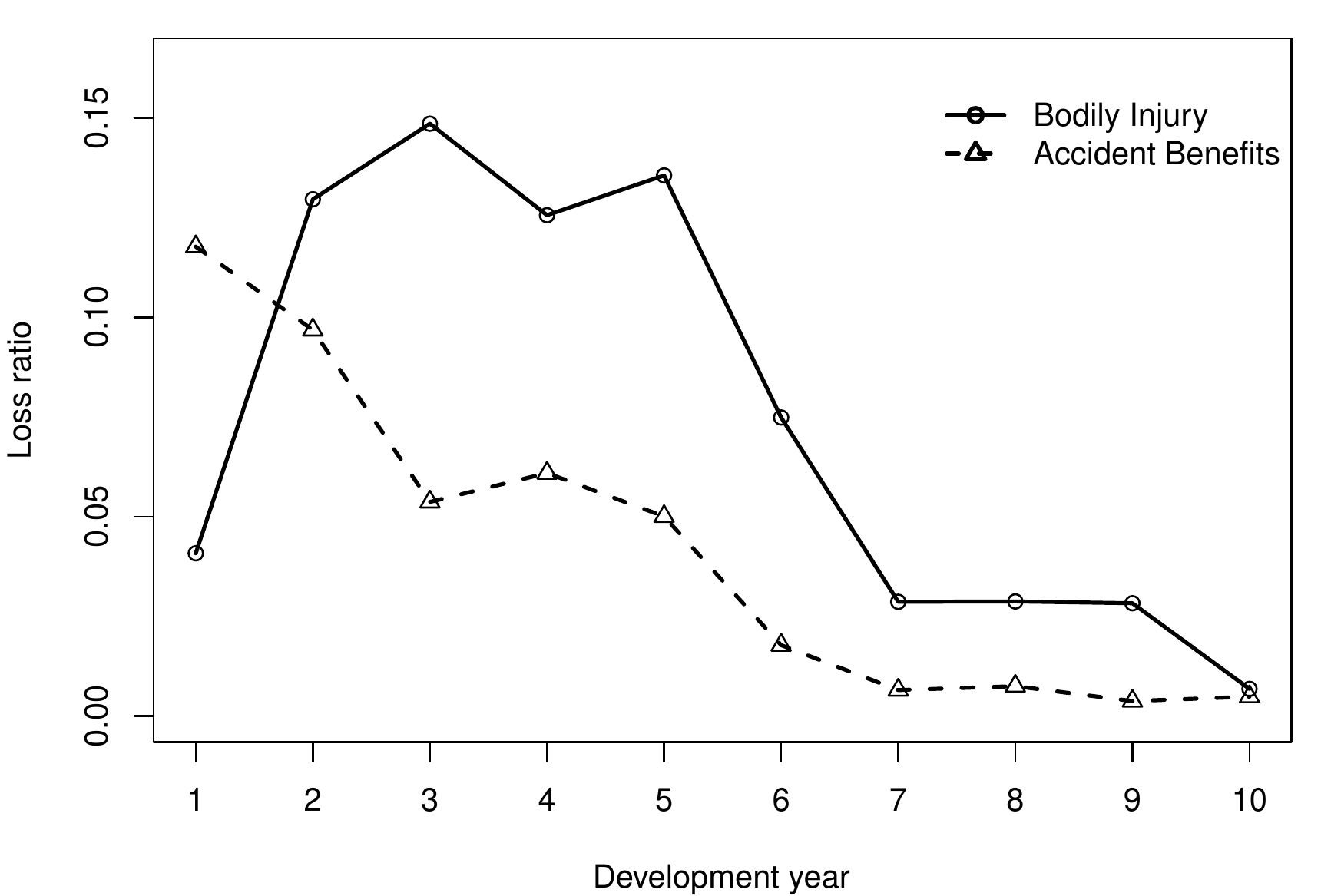}
		\subcaption{Plot of loss ratios for accident year 2003}
	\end{subfigure}
	\caption{(Colour online) Loss ratios from Bodily Injury line and Accident Benefits line (from a Canadian insurer)}\label{fig:unbalancedplot}
\end{figure}

{Figure \ref{fig:unbalancedplot} provides heat maps of incremental loss ratios on the left, and a plot of incremental loss ratios for accident year 2003 from the two lines of business on the right. For a given accident year, the loss ratio increment for development year $j$ is defined as the ratio of incremental claim payments in that development year to the earned premium for the accident year. Within a single loss triangle, one can observe a quite significant variation in claim observations across development years for any particular accident year. As shown in the heat map in Figure \ref{fig:unbalancedplot}, the claim activity for the Bodily Injury line is low in development year 0, then peaks in the next few years and dies out after the peak. This {typical} pattern is also shown in the plot of loss ratio on the right hand side of Figure \ref{fig:unbalancedplot} for accident year 2003. For the Accident Benefits line, the claim activity is the highest in development {year} 0 or 1, then drops quickly as we approach later development year. This pattern is also shown in the plot of loss ratios for accident year 2003 of the Accident Benefit line. The plot of ratios on the right hand side of Figure \ref{fig:unbalancedplot} also indicates the difference in development patterns for two different business lines. We can say that the Accident Benefits line is shorter-tailed than the Bodily Injury line. A variation can be observed across claim observations that come from the same accident year and the same development year, simply due to different claim development patterns across these business lines. This is in addition to the variation between loss values in different development lags and from different loss triangles, such as cells in development year 1 from the Bodily Injury line and cells in development year 10 from the Accident Benefit line.} \par 

{Overall, Figure \ref{fig:unbalancedplot} shows a large variation across claim observations in a loss reserving data set. Within a single loss triangle, there is variation across development years due to the development pattern of claims over time. Different claim development patterns can also result in variation between observations across loss triangles. Typically, one often does not expect dependence between lines that have different tail lengths, for example, an Auto Property Damage line is often independent of a Workers Compensation line. However, lines with different tail lengths can still have some association. One of such examples is a portfolio of Accident Benefits line and the Bodily Injury line in the above illustration.}\par 
	
{With the variations between claim observations within and across triangles, we refer to loss reserving data as unbalanced data. This data feature creates a number of challenges in applying common shock models to reserving data, which will be discussed in the remainder of this section. For generality and completeness, the focus is placed on the unbalanced feature of data consisting of multiple lines of business.}

\subsection{{General common shock framework}}
Consider $N$ loss triangles of claim cells $Y_{i,j}^{(n)}$. {The notation $Y_{i,j}^{(n)}$ can represent incremental claim payments or counts. We have the indices $i\, (i=1,...,I)$ representing} the accident period, $j\,(j=0,...,J)$ {representing} the development period, and $n\,(n=1,...,N)$ {representing} the business line.  It also follows that the claims $Y_{i,j}^{(n)}$ belong to the calendar period $t=i+j-1,\,(t=1,...,T)$. \par 

{Let $\mathcal{S}^{(n)} = \lbrace\mathcal{S}_s^{(n)};\,s=1,...,S\rbrace$ be a partition of the set of all claims $Y_{i,j}^{(n)}$ from business line $n$. Also assume that all partitions are the same for different lines $n$ for simplicity. Denote by $\boldsymbol{\pi}_{(i,j)}=\rev{s}$ a unique mapping of claim $Y_{i,j}^{(n)}$ to a set $\mathcal{S}_s^{(n)}$ in the partition. For example, the partition $\mathcal{S}^{(n)} = \lbrace\mathcal{S}_s^{(n)};\,s=1,...,I\rbrace$ where $\mathcal{S}_s^{(n)} = \lbrace Y_{s,j}^{(n)};\, j=1,...,J\rbrace$ represents a partition of claims by accident year. The selection of the partition $\mathcal{S}^{(n)}$ is very flexible and can be {specified} for different types of dependence.}\par  

Many multivariate models with different types of dependence can be generalised by the common shock framework in \citet*{AvTaWo18} with 
	\begin{equation}
	Y_{i,j}^{(n)} = \kappa_{i,j}^{(n)}W_{\boldsymbol{\pi}_{(i,j)}} + \lambda_{i,j}^{(n)}U_{\boldsymbol{\pi}_{(i,j)}}^{(n)} + Z_{i,j}^{(n)}, \label{Eq:commonshock}
	\end{equation}
{where $\boldsymbol{\pi}_{(i,j)}=s$ denote the unique mapping of the claim $Y_{i,j}^{(n)}$ to the corresponding subset $\mathcal{S}_s^{(n)}$ in the partition $\mathcal{S}^{(n)}$, and} $W_{\boldsymbol{\pi}_{(i,j)}},\,U_{\boldsymbol{\pi}_{(i,j)}}^{(n)},\,Z_{i,j}^{(n)}$ are independent stochastic variates. The common shock $W_{\boldsymbol{\pi}_{(i,j)}}$ introduces dependence {across all business lines $n=,1...,N$ on claims that belong to the {subsets} $\mathcal{S}_s^{(n)}$}. For example, accident year dependence across lines can be captured using the {partition where $\mathcal{S}_s^{(n)} = \lbrace Y_{s,j}^{(n)};\, j=1,...,J\rbrace$}. The other common shock $U_{\boldsymbol{\pi}_{(i,j)}}^{(n)}$ introduces dependence across claims within the set {$\mathcal{S}_s^{(n)}$} of business line $n$ only, such as development year dependence with {the partition set specification $\mathcal{S}_s^{(n)} =\lbrace Y_{i,s}^{(n)};\, i=1,...,I\rbrace$}. Overall, the flexibility of choice of the subsets  $\mathcal{S}_s^{(n)}$ allows different dependence structures to be captured. Lastly, the idiosyncratic component, which is unique to the claim $Y_{i,j}^{(n)}$, is denoted by $Z_{i,j}^{(n)}$. Scaling factors, denoted by  $\kappa_{i,j}^{(n)},\,{\lambda}_{i,j}^{(n)}$, control the extent to which the set-wide common shock contributes to individual members of the set. \rev{In this section, we have wished to preserve the link to the general notation of \citet{AvTaWo18} through the use of the notation $\boldsymbol{\pi}_{(i,j)}$. This notation will be simplified in Section \ref{sec:balanceprop} for specific examples.}\par 
\begin{remark}
	\rev{There can be situations where variables $\lbrace Y_{i,j}^{(n)}; \forall i,\,j;\,n> 2\rbrace$ are pairwise dependent (i.e. the dependency between each pair of variables is driven by a different source). \rev{For example, there can be a portfolio of 3 lines of business (LOBs) where there are 3 independent common shocks that drive the dependence between each of the following three pairs, LoB 1 and LoB 2, LoB 2 and LoB 3 and LoB 3 and LoB 1, respectively.} In such cases, one can consider having additional common shock variables $W_{\boldsymbol{\pi}_{(i,j)}}$ that capture dependence across lines, \rev{for example, $W_{\boldsymbol{\pi}_{(i,j)}}^{(1,2)}$, $W_{\boldsymbol{\pi}_{(i,j)}}^{(2,3)}$, $W_{\boldsymbol{\pi}_{(i,j)}}^{(3,1)}$ for the above scenario of 3 LoBs}. However, it is worth noting that these will result in more parameters required for the framework.}
\end{remark}

\subsection{Balancing common shock proportions in loss reserving data}\label{sec:balanceprop}
As a result of the unbalanced feature of reserving data, a common shock model can create problems in the absence of careful modelling. 

\rev{Consider a special case of Equation \eqref{Eq:commonshock} for dependence within a business line (i.e. $W_{\boldsymbol{\pi}_{(i,j)}}=0$). Further specify accident period dependence (i.e. $\boldsymbol{\pi}_{(i,j)}^{(n)} =  p$ for the mapping of subsets in the partition where $\mathcal{S}_s^{(n)} = \lbrace Y_{s,j}^{(n)};\, j=1,...,J\rbrace$). This allows us to simplify  $U_{\boldsymbol{\pi}_{(i,j)}}^{(n)}={X_{i}}^{(n)}$. Hence the general framework is reduced to} 
	\begin{align}
	Y_{i,j}^{(n)} = \lambda_{i,j}^{(n)}\rev{X_{i}^{(n)}} + Z_{i,j}^{(n)}. \label{Eq:commonshockwithin}
	\end{align}
	Consequently, the proportionate contribution of the common shock to the expected value of the total observation is 
	\begin{align}
	\frac{\lambda_{i,j}^{(n)}\text{E}\left[\rev{X_{i}^{(n)}}\right]}{ \lambda_{i,j}^{(n)}\text{E}\left[\rev{X_{i}^{(n)}}\right] + \text{E}\left[Z_{i,j}^{(n)}\right]}.
	\end{align}
If the scaling factor is removed, i.e. $\lambda_{i,j}^{(n)}=1$, this proportion has an inverse relationship with the mean of the idiosyncratic component $\text{E}\left[Z_{i,j}^{(n)}\right]$. As a result, in a set of loss cells in a triangle that are dependent and share a common shock, the cells with large values have a smaller proportion of common shock contribution and vice versa. This is because claims within the same accident period, or within the same calendar period belong to different development periods. As explained in Section \ref{Sec:unbalancednature}, their values can vary significantly due to the variation in claim activity across development periods. \rev{This issue can also be observed in the case of calendar period dependence (i.e. {\rev{$\boldsymbol{\pi}_{(i,j)}^{(n)} =  s$} for the mapping of subsets in the partition where {$\mathcal{S}_s^{(n)} = \lbrace Y_{i,s-i+1}^{(n)};\, i=1,...,J\rbrace$}}). }\par 
		 
A similar issue is encountered for a portfolio of dependent business lines with differing tail lengths{, such as the two business lines Bodily Injury and Accident Benefits in the illustration in Section \ref{Sec:unbalancednature}}. We consider a special case of Equation \eqref{Eq:commonshock} {{that} allows for} dependence between business lines only (i.e. ${U_{\boldsymbol{\pi}_{i,j}}^{(n)}=0}$). \rev{Further specify cell-wise dependence (i.e. partition mapping where $\mathcal{S}_{i,j}^{(n)} = \lbrace Y_{i,j}^{(n)}\rbrace$). This allows us to simplify  $W_{\boldsymbol{\pi}_{(i,j)}}={V_{i,j}}$. }The contribution of the common shock to the total expected observation is then given by 
\begin{align}
\frac{\kappa_{i,j}^{(n)}\text{E}\left[V_{i,j}\right]}{ \kappa_{i,j}^{(n)}\text{E}\left[V_{i,j}\right] + \text{E}\left[Z_{i,j}^{(n)}\right]}.
\end{align}
If the scaling factor is removed, i.e. $\kappa_{i,j}^{(n)}$, this proportion also has an inverse relationship with the mean of the idiosyncratic component $\text{E}\left[Z_{i,j}^{(n)}\right]$. As {explained} in Section \ref{Sec:unbalancednature}, values of claims in a portfolio of multiple triangles can vary in two main ways: across development years within a loss triangle, and across loss triangles. As a result, the proportion of common shock varies within and across loss triangles wherein loss cells with larger values have smaller common shock contributions. \rev{In the case of pairwise dependence considered above, the disproportion is typically a result of varying tail lengths \rev{across} business lines. However, it is worth noting that unbalanced common shock proportions can also be typically observed for accident year dependence, or calendar dependence across business lines \rev{from the same cause}.}\par
	
\begin{figure}[htb]
	\centering
		\includegraphics[scale=0.67]{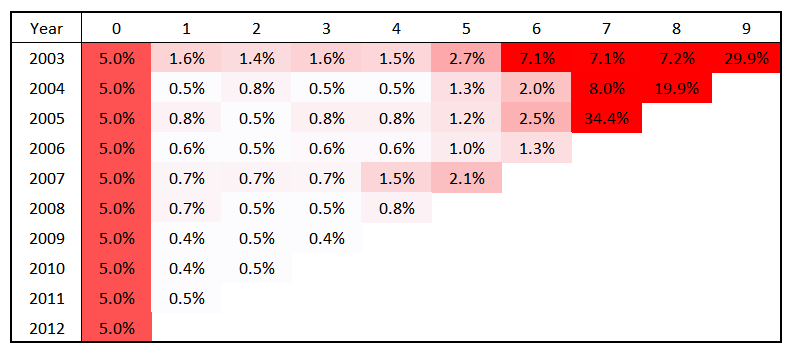}\\
		\includegraphics[scale=0.67]{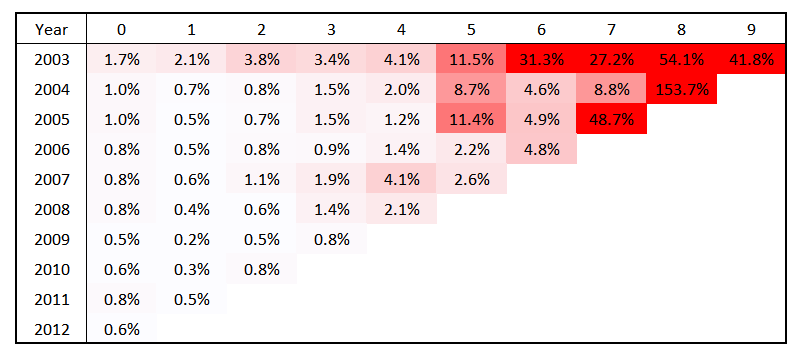}
		\caption{(Colour online) Heat maps of common shock contributions in Bodily Injury line (top) and Accident Benefit line (bottom) without using scaling terms}\label{fig:prop_unbalanced}
\end{figure}

{We consider the case of accident year dependence across the two triangles illustrated in Section \ref{Sec:unbalancednature} \rev{(i.e. partition mapping $\mathcal{S}_s = \lbrace Y_{s,j}^{(n)};\, j=1,...,J; n=1,...,N\rbrace$, and we can simplify $W_{\boldsymbol{\pi}_{(i,j)}}={V_{i}}$)}. For illustration, the mean of the common shock $E\left[\rev{V_{i}}\right]$ is set to 5\% of the {loss ratios} in the first development year of each accident year in the Bodily Injury line. The contributions of common shock are shown in Figure \ref{fig:prop_unbalanced} assuming no scaling terms. With accident year dependence across business lines, claims within the same accident year share the same common shock. These include claims from different development years within and across loss triangles. Because their values vary due to different claim activities within and across lines, their common shock proportions also vary. Specifically, common shock proportions are significantly {smaller} in areas with high claim activity, and {larger} in areas with low claim activity, as shown in Figure \ref{fig:prop_unbalanced}.} 

     In general, quite significant variations in common shock proportions can be observed within and across segments in the absence of careful modelling as a result of the unbalanced nature of loss reserving data. One may wish to confine the relation of the common shock to total observations over the entire range of the triangles. 
	
	\subsection{Maintaining model parsimony}\label{Sec:Parsimony}
	 The most {straight-forward} solution to the balancing common shock proportions within and across triangles is to have cell-specific scaling factors $\kappa_{i,j}^{(n)},\,\lambda_{i,j}^{(n)}$ to adjust the common shock effects for each total observation $Y_{i,j}^{(n)}$. However, this implies that $2IJN$ new parameters are required for the entire range of triangles of observed data and outstanding claims to be predicted. Given that the variation in claim observations typically occurs across development periods, one may simplify the scaling factors to be column-specific $\kappa_{i,j}^{(n)}=\kappa_{j}^{(n)},\,\lambda_{i,j}^{(n)}=\lambda_{j}^{(n)}$. However, this still results in $2JN$ new parameters. \par 
		
 Loss triangle data typically has a small sample size. While the presence of scaling factors can mitigate the impact of the unbalanced nature of reserving data, it also adds many more parameters to the model. If scaling factors are not chosen carefully, it may result in over-fitting and the number of parameters to be estimated can even exceed the number of observations.
	
	\subsection{Maintaining distributional tractability}\label{Sec:Tractability}
     On some occasions, parameters $\lambda_{i,j}^{(n)}$ and $\kappa_{i,j}^{(n)}$ need to be specified such that the total observation $Y_{i,j}^{(n)}$ follows a specific distribution \citep*{AvTaWo18}. This is referred to as distributional tractability, or closure under the taking of marginals, which is considered in \citet*[Chapter 4]{Joe97} to be one of the four desirable properties of a multivariate model (see also Section \ref{Sec:Intro}). \par 
    
     Consider as an example the common shock Tweedie framework in \citet*{AvTaVuWo16}. This framework is developed for cell-wise dependence across business lines  (i.e. {$\mathcal{S}_{i,j}^{(n)} = \lbrace Y_{i,j}^{(n)}\rbrace$, $U_{i,j}^{(n)}=0$}). Fitting this into the general common shock structure in Equation \eqref{Eq:commonshock} {and simplifying $W_{\boldsymbol{\pi}_{(i,j)}}={V_{i,j}}$} we have 
    \begin{align}
    Y_{i,j}^{(n)} & =   
    \kappa_{i,j}^{(n)}{V_{i,j}}+Z_{i,j}^{(n)},\label{Eq:commonshockTweedie}
    \end{align}
    where the two components ${V_{i,j}},\,Z_{i,j}^{(n)}$ are assumed to be independent and have Tweedie distributions
    \begin{align}
    {V_{i,j}} & \sim   \text{Tweedie}_p(\alpha,\beta),  \label{E_Wij} \\
    Z_{i,j}^{(n)} & \sim  \text{Tweedie}_p(\eta_i^{(n)}\nu_j^{(n)}, \gamma^{(n)}). \label{zmarginalreprod}
    \end{align}
     Parameter $p$ is the power parameter which specifies a member of the Tweedie family, for example $p=1$ corresponds to a Poisson distribution. The representation of Tweedie distributions used is the reproductive representation \citep*[Chapter 4]{Jor97}. This representation specifies a Tweedie random variable using a location (or mean) parameter, and a dispersion parameter. In the above model specification, parameters $\alpha$ and $\eta_i^{(n)}\nu_j^{(n)}$ are the location parameters, and parameters $\beta$ and $\gamma^{(n)}$  are the dispersion parameters. The reproductive representation has a distinctive property wherein the weighted average of independent Tweedie variables with the same power parameter $p$ and the same location parameter is also a Tweedie variable with the same power and location parameters. The weighting factors are determined using dispersion parameters of the component variables in the weighted average.\par
     		
     It then follows that the mean and variance of the two components ${V_{i,j}},\,Z_{i,j}^{(n)}$ are
     \begin{align}
    \text{E}\left[{V_{i,j}}\right]   &=  \alpha, & \qquad  \text{Var}\left[{V_{i,j}}\right] &= \beta\alpha^p,\\
     \text{E}\left[Z_{i,j}^{(n)}\right]  &= \eta_i^{(n)}\nu_j^{(n)}, & \qquad  \text{Var}\left[Z_{i,j}^{(n)}\right]  &=  \gamma^{(n)}(\eta_i^{(n)}\nu_j^{(n)})^p.
    \end{align}
    As stated in Remark 2.2 of \citet*{AvTaVuWo16}, the most simple parametrisation is used for the common shock component ${V_{i,j}}$ with parameters $\alpha$ and $\beta$. \par 
    
    \rev{As mentioned earlier in this section, it can be desirable to maintain distributional tractability, or closure under the taking of marginals for ease of interpretation.} It follows from the form of closure under addition of the Tweedie family of distributions, as proven in \citet*[Chapter 3]{Jor97}, that a specific choice of $\kappa_{i,j}^{(n)}$ is required to ensure \rev{that} $Y_{i,j}^{(n)}$ \rev{also has a Tweedie distribution}. This choice is
    \begin{align}
    \kappa_{i,j}^{(n)} = \left(\dfrac{\alpha}{ \eta_i^{(n)}\nu_j^{(n)} }\right)^{1-p}\dfrac{\gamma^{(n)}}{\beta}.\label{Eq:commonshockcoef}
    \end{align}
    \par 
 
    The mean expression is given by 
    \begin{equation}
    \text{E}\left[Y_{i,j}^{(n)}\right] =  \left(\dfrac{\alpha}{\eta_i^{(n)}\nu_j^{(n)}}\right)^{2-p}\dfrac{\gamma^{(n)}}{\beta} \eta_i^{(n)}\nu_j^{(n)} +\eta_i^{(n)}\nu_j^{(n)}
    \end{equation}  
    where the first term in the summation is the contribution from the common shock and the second term is the contribution from the idiosyncratic component. The expected contribution of the common shock to the total expected observation is
    \begin{equation}
    \dfrac{\left(\dfrac{\alpha}{\eta_i^{(n)}\nu_j^{(n)}}\right)^{2-p}\dfrac{\gamma^{(n)}}{\beta} }{\left(\dfrac{\alpha}{\eta_i^{(n)}\nu_j^{(n)}}\right)^{2-p}\dfrac{\gamma^{(n)}}{\beta}  +1}. \label{commonshockprop}
    \end{equation}
    The following observation can be made on the effect of the power parameter $p$:
    \begin{arcitem}
    	\item  If $p<2$:  The above ratio increases as $\nu_j^{(n)}$ decreases. As a result, the proportion of common shock is understated in early development periods, and overstated in late development periods \citep*{AvTaWo18}. In a portfolio of segments with varying tail lengths, the larger the discrepancy between the tail lengths (i.e. between $\nu_j^{(n)}$ and $\rev{\nu_j^{(m)}}$), the larger the variation in the common shock contributions. \rev{The behaviour of the above ratio has been examined with respect to development factor $\nu_j^{(n)}$ in particular because variation within and across lines of business is mainly driven by the development pattern of claims as explained in Section \ref{Sec:unbalancednature}. As a result, one would expect the development factors to vary the most.}
    	\item If $p>2$: The opposite observation is made for the relationship between the above ratio and $\nu_j^{(n)}$ \rev{(i.e. the above ratio decreases as $\nu_j^{(n)}$ decreases).}
    	\item If $p=2$: In this special case, the common shock contribution is simplified to 
    	\begin{equation}
    \dfrac{\dfrac{\gamma^{(n)}}{\beta}}{\dfrac{\gamma^{(n)}}{\beta}+1},
    	\end{equation}
    	 which is now independent of accident and development periods. Consequently, the common shock contributes proportionately to the total observations over the entire range of the triangles. It is also worth emphasising that specifying $p=2$ gives the multivariate gamma case of the multivariate Tweedie framework.
    \end{arcitem} 
    
     The above analyses and examples show that the choices of scaling factors $\kappa_{i,j}^{(n)}$  and $\lambda_{i,j}^{(n)}$  are subject to many constraints. To accurately capture the dependence structure, these  parameters are required to balance the common shock proportions within all claim observations over the entire range of the triangles. However, this can result in over-fitting, which can be a critical issue in loss reserving due to small sample size data. Furthermore, the specification of these parameters may need to be restricted in some cases for the purpose of preserving distributional tractability. It is then the aim of this paper to find a solution that  compromises between these conflicting issues with a specific application on the common shock Tweedie approach in \citet*{AvTaVuWo16}.
    
	\section{A common shock Tweedie approach to unbalanced data}\label{Sec:Model}
	 In this section, we propose a solution that compromises between conflicting challenges encountered by common shock models when they are applied to reserving data due to the unbalanced feature of the data. The focus of this development is on a common shock Tweedie approach to unbalanced data. The estimation method for this approach is also given. \par
	 
	 The multivariate Tweedie framework {described in Section \ref{Sec:Tractability}} is a typical example of an application of the common shock approach in stochastic loss reserving. It is of particular interest due to its various advantages. Developed on the Tweedie family of distributions, it offers flexible choices of marginal density that also include Tweedie's compound Poisson density with the ability to deal with zero data points. The framework can also be generalised to more than two dimensions. In addition, the explicit common shock structure allows the correlation matrix to be obtained in closed form. Moment- and cumulant-generating-functions can also be obtained analytically, enhancing the tractability of the model. Similar to other common shock models, this framework also encounters the issue of unbalanced data. As explained in Section \ref{Sec:Challenges}, the selection of scaling coefficients for the common shock term in this framework is constrained by the need to balance common shock proportions while maintaining model parsimony and distributional tractability. 
	\subsection{{Theoretical framework}}
	Claims are first standardised using a common unit of exposure such as the number of claims, or the total amount of premium collected, to ensure consistency across accident periods and business lines. {Recall the specification of the common shock Tweedie model in \citet{AvTaVuWo16} described in Section \ref{Sec:Tractability},}
		\begin{align}
		{Y_{i,j}^{(n)}} & {=   
			\kappa_{i,j}^{(n)}V_{i,j}+Z_{i,j}^{(n)},}\tag{\ref{Eq:commonshockTweedie}}
		\end{align}
		where
			\begin{align}
	{V_{i,j}} & {\sim   \text{Tweedie}_p(\alpha,\beta),  \tag{\ref{E_Wij}}} \\
	{Z_{i,j}^{(n)}} & {\sim  \text{Tweedie}_p(\eta_i^{(n)}\nu_j^{(n)}, \gamma^{(n)}) \tag{\ref{zmarginalreprod}}},\\
	  {\kappa_{i,j}^{(n)}} &= {\left(\dfrac{\alpha}{ \eta_i^{(n)}\nu_j^{(n)} }\right)^{1-p}\dfrac{\gamma^{(n)}}{\beta}.\tag{\ref{Eq:commonshockcoef}}}
	\end{align}
	{Recall that $\alpha$ and $\eta_i^{(n)}\nu_j^{(n)}$ are location (mean) parameters, and $\beta$ and $\gamma^{(n)}$ are dispersion parameters of ${V_{i,j}}$ and $Z_{i,j}^{(n)}$ respectively.}
	
	{As shown in Equation \eqref{Eq:commonshockcoef}, the common shock scaling factor has to be specified in the above form that involves parameters of the common shock $V_{i,j}$ and the idiosyncratic component $Z_{i,j}^{(n)}$. However, due to the unbalanced feature of reserving data with $\nu_j^{(n)}$ {varying} across development lag $j$ and business line $n$, the common shock contributes disproportionately to the total observation $Y_{i,j}^{(n)}$. It is also desirable to maintain model parsimony.}\par
	
	{Given the above considerations, we can replace the non-cell-specific parameter $\alpha$ in the scaling factor with column-specific parameter}
	\begin{align}
	{{\alpha_j}} &{=\rev{\tilde{c}}	\left(\prod_{n}\text{E}\left[Z_{i,j}^{(n)}\right]\right)^{\frac{1}{N}} = \rev{\tilde{c}}\left(\prod_{n}{\eta}_i^{(n)}{\nu}_j^{(n)}\right)^{\frac{1}{N}}} \\
	 &{\approx c\sqrt[N]{\nu_j^{(1)}...\nu_j^{(N)}}}.
	\end{align}
	{The parameter $\alpha_j$ is also the location parameter of the common shock ${V_{i,j}}$. As a result, we \rev{approximately} have} 
		\begin{align}
		{V_{i,j}} & {\sim   \text{Tweedie}_p({\alpha_j},\beta) =  { \text{Tweedie}_p\left(c\sqrt[N]{\nu_j^{(1)}...\nu_j^{(N)}},\beta\right)}. \label{Eq:shock}}
		\end{align}
		\par 
	{Essentially, the common shock parameter $\alpha_j$ {is} proportional to the geometric average of idiosyncratic components of claims which share the same common shock component. In this case, these are claims in the same accident period and development period as the framework is used to capture cell-wise dependence. This geometric average can then be simplified by removing accident period factors because we can reasonably expect limited variation across accident periods as a result of claims standardisation, assuming no significant changes occur across accident periods.}\par 
	
	{The above specification of scaling factor aims to balance the impact of unbalanced feature in loss reserving data which is mainly introduced by variations in development factors $\nu_j^{(n)}$. Using this specification, the common shock proportion is given by}
	\begin{equation}
	{\dfrac{\left(\dfrac{c\sqrt[N]{\nu_j^{(1)}...\nu_j^{(N)}}}{\eta_i^{(n)}\nu_j^{(n)}}\right)^{2-p}\dfrac{\gamma^{(n)}}{\beta} }{\left(\dfrac{c\sqrt[N]{\nu_j^{(1)}...\nu_j^{(N)}}}{\eta_i^{(n)}\nu_j^{(n)}}\right)^{2-p}\dfrac{\gamma^{(n)}}{\beta}  +1}. } \label{commonshockprop3}
	\end{equation}
	{This does not provide a complete balance of common shock proportions because the effect of $\nu_j^{(n)}$ is reduced by a factor $\sqrt[N]{\nu_j^{(n)}}$. However it still provides quite a significant improvement over the original framework. This will be demonstrated in the simulation illustration in Section \ref{Sec:sim}. This specification can also preserve distributional tractability of the framework. In addition, model parsimony is retained as the total number of parameters in the framework is unchanged. This can be considered an {effective} solution given the three constraints discussed in Section \ref{Sec:Challenges}.}
	
	{In addition to the above treatment for unbalanced data, we also introduce a treatment for negative claims}
	\begin{align}
	Y_{i,j}^{(n)} + \xi^{(n)}  & = 
	\left(\dfrac{ {c\sqrt[N]{\nu_j^{(1)}...\nu_j^{(N)}}}}{ \eta_i^{(n)}\nu_j^{(n)} }\right)^{1-p}\dfrac{\gamma^{(n)}}{\beta}{V_{i,j}}+Z_{i,j}^{(n)},\label{Eq:modelstructure}
	\end{align}
	where a translation factor is used and defined such that
	\begin{align}
	\xi^{(n)} & = \begin{cases}
	0 & \text{if } {\min\lbrace Y_{i,j}^{(n)},\,\forall i,\,j \rbrace\ge 0},\\
	\ge -\min\lbrace Y_{i,j}^{(n)}\rbrace & \text{if } {\min\lbrace Y_{i,j}^{(n)},\,\forall i,\,j\rbrace < 0}. \\
	\end{cases}\label{Eq:scalingfactor}
	\end{align}
	 The translation is only needed for a loss triangle if it contains at least one negative value and it must be large enough to offset the smallest negative value. It is worth emphasising that in this case, while its lower bound is deterministic, the actual value of $\xi^{(n)}$ still has to be estimated. The generalisation of this treatment to the general common shock framework in \citet*{AvTaWo18} is straightforward. 

	{Following from the {above specification},} the marginal density is then given by
	\begin{align}
	Y_{i,j}^{(n)} + \xi^{(n)} \sim 
	\text{Tweedie}_p\left( \eta_i^{(n)}\nu_j^{(n)} \left[\left(\dfrac{ {c\sqrt[N]{\nu_j^{(1)}...\nu_j^{(N)}}}}{ \eta_i^{(n)}\nu_j^{(n)} }\right)^{2-p} \dfrac{\gamma^{(n)}}{\beta}+1\right],\gamma^{(n)}\left[\left(\dfrac{ {c\sqrt[N]{\nu_j^{(1)}...\nu_j^{(N)}}}}{ \eta_i^{(n)}\nu_j^{(n)} }\right)^{2-p} \dfrac{\gamma^{(n)}}{\beta}+1\right]^{1-p}\right), \label{Eq:marginal}
	\end{align}
	where the first parameter is the location parameter and also the mean of $Y_{i,j}^{(n)} + \xi^{(n)}$. The second parameter is the dispersion parameter. It follows that the vector of translated claims in the same position across all triangles
	\begin{align}
	 {{}_{\xi}\boldsymbol{Y}_{i,j} =  \begin{pmatrix}
	{Y}_{i,j}^{(1)} + {\xi}^{(1)}\\
	{Y}_{i,j}^{(2)} + {\xi}^{(2)}\\
	\vdots\\
	{Y}_{i,j}^{(N)} + {\xi}^{(N)}\\
	\end{pmatrix},}
	\end{align}
	has a multivariate Tweedie distribution with the multivariate density
	\begin{align}
	& f_{{}_{\xi}\boldsymbol{Y}_{i,j}}\left(y_{i,j}^{(1)}+\xi^{(1)},...,y_{i,j}^{(N)}+\xi^{(N)}\right) =\int_{0}^{A_{i,j}} f_{{V_{i,j}}}(w_{i,j})\prod_{n=1}^{N}f_{Z_{i,j}^{(n)}}\left(y_{i,j}^{(n)} + \xi^{(n)}-\left(\dfrac{{ {c\sqrt[N]{\nu_j^{(1)}...\nu_j^{(N)}}}}}{ \eta_i^{(n)}\nu_j^{(n)} }\right)^{1-p}\dfrac{\gamma^{(n)}}{\beta}w_{i,j}\right)dw_{i,j},\label{Eq:multivariate}
	\end{align}
	where
	\begin{equation}
	A_{i,j} = {\min \left(\left(\dfrac{ \eta_i^{(1)}\nu_j^{(1)} }{ {c\sqrt[N]{\nu_j^{(1)}...\nu_j^{(N)}}}}\right)^{1-p}\dfrac{\beta}{\gamma^{(1)}}(y_{i,j}^{(1)}+\xi^{(1)}),...,\left(\dfrac{ \eta_i^{(N)}\nu_j^{(N)} }{ {c\sqrt[N]{\nu_j^{(1)}...\nu_j^{(N)}}}}\right)^{1-p}\dfrac{\beta}{\gamma^{(N)}}(y_{i,j}^{(N)}+\xi^{(N)})\right)},
	\end{equation}
	and where $f(.)$ is the Tweedie density in reproductive form \citep[see also][Chapter 4]{Jor97}. 
	
	\subsection{Model estimation with Bayesian inference}\label{Sec:est}
	Bayesian inference is used for model estimation. Bayesian estimation has gained its popularity in the loss reserving literature due to rapid computing advancements and Markov Chain Monte Carlo (MCMC) methods that allow the calculation of intractable posterior densities to be performed significantly faster \citep*{AvTaVuWo16,VeHoBj12}. In addition, the incorporation of prior densities in the calculation of posterior densities is a natural way to allow for parameter error in modelling \citep*{ShBaMe12,EnVeWu12}. Another aim of using a Bayesian set-up is to also estimate the power parameter $p$ and translation parameter $\xi^{(n)}$ with allowance of parameter uncertainty. This is to formalise the estimation of these parameters as they are often estimated heuristically in practice. {It is worth emphasising that the Bayesian structure is not integral to our model,  but serves as a device for estimation.}
	
	 A two step procedure is used for estimation, similar to that in \citet*{AvTaVuWo16}. The first stage is the estimation of all parameters except $c$ and $\beta$ of the common shock ${{V_{i,j}}}$. This stage, however, gives the estimate of {a ratio of these parameters denoted as} 
	 \begin{align}
	 \delta = \frac{c^{2-p}}{\beta},
	 \end{align}
	 as can observed from Equation \eqref{Eq:marginal}.  This is followed by the multivariate stage that estimates $c$ and $\beta$ conditional on estimates of other parameters from the first stage. The motivation for this procedure comes from properties of the common shock Tweedie framework.  Claim observations in the same position across triangles in this framework follow a multivariate Tweedie distribution, and each observation itself also has a marginal Tweedie distribution. In addition, the multivariate density has an integral calculation, as shown in Equation \eqref{Eq:multivariate}. This can prolong the estimation of the posterior density, making the tuning and convergence of MCMC much more difficult.
	
	A Bayesian set-up requires the specification of the likelihood functions, prior densities and, if posterior densities are not in closed form, computational algorithms used to approximate them. The likelihood functions follow from  Equation \eqref{Eq:marginal} for the first stage and Equation \eqref{Eq:multivariate} for the second stage. 
	
	Prior densities are then specified. Prior densities can be chosen to be informative or uninformative. Uninformative priors assign equal possibilities to all values in the feasible set of parameter values, whereas informative priors convey some prior preference for certain values of the parameters. However, the use of informative priors can significantly improve the convergence rate, especially when the parameter dimension is large \citep*{Con10}. Parameter estimates from univariate Tweedie model \citep*{AlWu09} can assist in the specification of informative prior densities for parameters $\eta_{i}^{(n)}$, $\nu_j^{(n)}$ and $\gamma^{(n)}$. A preliminary analysis of the dependence structure can help select informative prior densities for the common shock parameters $c$ and $\beta$. Regarding  {the} prior densities for $p$ and $\xi^{(n)}$, some  {constraints} need to be taken into account. In particular, $p$ is not defined in $(0,1)$, and $\xi^{(n)}$ has a lower bound as per its specification in Equation \eqref{Eq:scalingfactor}.
	
	Putting together the likelihood and prior specifications, the posterior density in the first stage is given by 
	\begin{equation}
	f_{\boldsymbol{\Omega}|\boldsymbol{Y^U}}(\boldsymbol{\Omega}|\boldsymbol{Y^U}) \propto \left(\prod_{i,j,n}f_{Y_{i,j}^{(n)}+ {\xi^{(n)}}}\left({y_{i,j}^{(n)}}+ {\xi^{(n)}}|\boldsymbol{\Omega}\right)\right)f_{p}(p)f_{\boldsymbol{\xi}}(\boldsymbol{\xi})f_{\delta}(\delta)f_{\boldsymbol{ {\eta}}}(\boldsymbol{ {\eta}})f_{\boldsymbol{ {\nu}}}(\boldsymbol{ {\nu}})f_{\boldsymbol{\gamma}}(\boldsymbol{\gamma}),\label{Eq:marginalpost}
	\end{equation}
	where
	\begin{gather*}
	\boldsymbol{\Omega} = \begin{pmatrix}
	p\\
	\boldsymbol{\xi}\\
	\delta\\
	\boldsymbol{ {\eta}} \\
	\boldsymbol{ {\nu}}\\
	\boldsymbol{\gamma}\\
	\end{pmatrix}, \,\boldsymbol{\xi} = \begin{pmatrix}
	\xi^{(1)}\\
	\xi^{(2)}\\
	\vdots\\
	\xi^{(N)}
	\end{pmatrix},\,\boldsymbol{\eta_i} = \begin{pmatrix}
	\eta_i^{(1)}\\
	\eta_i^{(2)}\\
	\vdots\\
	\eta_i^{(N)}
	\end{pmatrix},\, {\boldsymbol{\eta} = \begin{pmatrix}
	\boldsymbol{\eta_2}\\
	\boldsymbol{\eta_3}\\
	\vdots\\
	\boldsymbol{\eta_I}
	\end{pmatrix}},\, \boldsymbol{\nu_j} = \begin{pmatrix}
	\nu_j^{(1)}\\
	\nu_j^{(2)}\\
	\vdots\\
	\nu_j^{(N)}
	\end{pmatrix},\,  {\boldsymbol{\nu} = \begin{pmatrix}
	\boldsymbol{\nu_1}\\
	\boldsymbol{\nu_2}\\
	\vdots\\
	\boldsymbol{\nu_J}
	\end{pmatrix}},\,\boldsymbol{\gamma} = \begin{pmatrix}
	\gamma^{(1)}\\
	\gamma^{(2)}\\
	\vdots\\
	\gamma^{(N)}
	\end{pmatrix},
	\end{gather*}
	and where $\boldsymbol{Y^U}$ is a vector of claim observations \rev{in the upper claim triangles}. \par 
	
	From the model structure in Equation \eqref{Eq:modelstructure}, we have that all claims $Y_{i,j}^{(n)}$ are independent conditional on common shock. Hence, the joint likelihood can be written as a product of two separate parts: a product of the densities of claims conditional on common shock, and the density of the common shock. {In the first stage of the estimation procedure}, the likelihood obtained is the first part of the joint likelihood. {As also mentioned earlier, this stage provides the estimates of mean parameters $\boldsymbol{\nu},\,\boldsymbol{\eta}$ and dispersion parameters $\boldsymbol{\gamma}$ of the idiosyncratic variables $Z_{i,j}^{(n)}$, translation parameters $\boldsymbol{\xi}$ and power parameter $p$. This stage also provides the estimate of $\delta$ which is a function of parameters $c$ and $\beta$ of the common shock $V_{i,j}$.} \par 
	 
	 {In the second estimation step, we work with the joint likelihood directly since common shock components are not observed. In this step, the estimation of $c$ and $\beta$ is carried out conditioning on estimates of other parameters in the first step, including $\delta$ which is a function of $c$ and $\beta$. The multivariate Tweedie density of $\boldsymbol{Y}_{i,j}$ is used to obtain the likelihood in this estimation. The posterior density in this step} is given by 
	\begin{equation}
	f_{c|\boldsymbol{Y^U,\Omega}}({c}|\boldsymbol{Y^U,\Omega}) \propto \left(\prod_{i,j}f_{\boldsymbol{\boldsymbol{{}_\xi Y}_{i,j}}}\left(\boldsymbol{{}_\xi y}_{i,j}|c,\boldsymbol{\Omega}\right)\right)f_{c}(c).\label{Eq:multivariatepost}
	\end{equation}
	\par 
	
	The posterior densities in both stages are not in recognisable forms, hence MCMC algorithms are required for the estimation. The MCMC algorithm used is Metropolis-Hastings, which is a popular class of MCMC algorithms when the posterior distribution is not in a recognisable form. Random walk Metropolis-Hastings algorithms are used for marginal estimation and multivariate estimation. Proposal densities are chosen (tuned) so that acceptance probabilities are within desirable ranges. The tuning process can be done manually using classical Metropolis-Hastings algorithms. Alternatively, it can be done automatically in adaptive Metropolis-Hastings algorithms using coerced acceptance rates \citep*{HaSaTa01,Vih12}.

	\section{Simulation illustrations}\label{Sec:sim}
    Two illustrations are performed on two data sets. The first illustration, provided in Section \ref{Sec:sim1}, is to assess the {accuracy of the} estimation procedure. {Since true parameter values are known in a simulated data, a comparison of their estimates with their true values gives an indication of the appropriateness of the estimation procedure.} The second illustration, provided in Section \ref{Sec:sim2}, is to compare the performance of the common shock Tweedie approach with treatment for unbalanced data and the original common shock Tweedie approach in \citet*{AvTaVuWo16}. {This comparison focuses particularly on the contributions of common shock estimated from the two approaches.}
	
	\subsection{An illustration with unbalanced data and negative claims}\label{Sec:sim1}
	A data set consisting of two business lines, one of which has a negative claim observation, is simulated. The two loss triangles are represented in Tables \ref{Tab:sim1} and \ref{Tab:sim2} in Appendix \ref{Sec:data1}. {These two triangles {consist} of simulated claim observations.  Each observation in the triangles is drawn from the multivariate Tweedie model for unbalanced data represented in Section \ref{Sec:Model}. {For simplicity, these observations are assumed to have been adjusted for changes in exposure across accident years}.}
	
	The marginal fitting is first performed. Parameters are first transformed using the log transformation, and uniform prior densities are used. 200,000 simulations are run and 100,000 simulations are discarded as the burn-in period.  The sample chain is thinned by accepting every 5th iteration to reduce the serial dependence between iterations. MCMC paths of some parameters are given in Figure \ref{Fig:mcmc1}. A similar procedure is performed for the multivariate estimate. The estimates of $c$ and $\beta$ are obtained from this step. Parameter estimates are provided in Table \ref{Tab:marginal1} in Appendix \ref{Sec:data1}.\par
	
	{To evaluate the Bayesian inference used for estimation, we compare the true parameter values with 90\% confidence intervals obtained from the posterior distributions of these parameters. The results the true values always lie within the corresponding 90\% confidence intervals. This indicates the accuracy of the estimation procedure. }\par 
	
	\rev{We have calibrated the  model on the same simulated data set using sub-triangles of dimension $5\times 5$ to assess the robustness of the proposed calibration method. The results show that true values also fall within the 90\% confidence intervals of the estimates from this calibration. However, confidence intervals are generally larger than those from the calibration that uses full size triangles. This is expected due to higher uncertainty in the estimates coming from smaller sample size.}
	
	\rev{To evaluate the bias in the resulting reserve predictions, forecasts of outstanding claims using our model are compared with the true forecasts as well as forecasts from a multivariate chain ladder model. The true forecasts are calculated as the expected value of outstanding claims using true parameter values. The multivariate chain ladder model used is the model in \citet{PrSc05}. The results are shown in Table \ref{tab:simdatacompare}.} 
		\begin{table}[H]
			\centering
			\begin{tabular}{ccccccc}
				\toprule
				\hline
				\multirow{2}{*}{{LoB}} & \multicolumn{3}{c}{{Balanced Multivariate Tweedie}} &  \multirow{2}{*}{{True forecasts}} & \multicolumn{2}{c}{{Multivariate Chain Ladder}}\\
				&{Mean} & {Standard error}  & {90\% CI} &&{Mean} & {Standard error}\\
				\midrule
				{1} & {159.38} & {19.74} &  {(129.64;193.88)}& {157.56} & {138.97} & {16.55}\\
				{2} & {598.98} & {76.15} & {(484.99;734.18)} & {563.92} & {531.31} & {52.17} \\
				{Total} &  {758.37} &{82.88} & {(633.14; 904.17)} & {721.48} &{670.28}& {61.31} \\
				\hline
				\bottomrule
			\end{tabular}
			\caption{\rev{Comparison of outstanding claims forecasts}}\label{tab:simdatacompare}
		\end{table} 
		\rev{It can be observed from Table \ref{tab:simdatacompare} that the true forecasts fall within the 90\% confidence intervals of the balanced multivariate Tweedie model forecasts. The forecasts from our model are also closer to the true forecasts than those from the multivariate chain ladder model. However, it is worth noting that the simulated data was generated from the multivariate Tweedie model in this illustration.} \par 
		
		\rev{We further assess bias in the resulting dependence structure by comparing the true cell-wise Pearson correlation coefficients for the outstanding claims and the cell-wise Pearson correlation coefficients calculated using the parameter estimates. Residual ratios, defined as ratios of estimated Pearson correlation coefficients to true Pearson correlation coefficients, are provided in Table \ref{tab:corrcompare}. The ratios are close to 1, indicating that the cell-wise dependence in the data is well captured.} 
		\begin{table}[H]
			\centering
			\begin{tabular}{cc|cccccccccc}
				\toprule
				\hline
				& & \multicolumn{10}{c}{Development year}\\
				&	& 1 & 2 & 3 & 4 & 5 & 6 & 7 & 8 & 9 & 10 \\
				\hline 
				\multirow{10}{*}{\rotatebox[origin=c]{90}{Accident year}}&
				1 &  &  &  &  &  &  &  &  &  &  \\ 
				&2 &  &  &  &  &  &  &  &  &  & 1.07 \\ 
				&3 &  &  &  &  &  &  &  &  & 0.99 & 0.99 \\ 
				&4 &  &  &  &  &  &  &  & 1.02 & 1.04 & 1.04 \\ 
				&5 &  &  &  &  &  &  & 1.00 & 0.98 & 0.99 & 1.00 \\ 
				&6 &  &  &  &  &  & 1.03 & 1.02 & 1.01 & 1.02 & 1.03 \\ 
				&7 &  &  &  &  & 1.02 & 1.03 & 1.03 & 1.01 & 1.03 & 1.03 \\ 
				&8 &  &  &  & 1.01 & 1.02 & 1.03 & 1.03 & 1.01 & 1.03 & 1.03 \\ 
				&9 &  &  & 1.04 & 1.03 & 1.04 & 1.05 & 1.05 & 1.03 & 1.05 & 1.05 \\ 
				&10 &  & 1.00 & 0.99 & 0.98 & 0.99 & 1.00 & 1.00 & 0.98 & 1.00 & 1.00 \\ 
				\hline
				\bottomrule
			\end{tabular}
			\caption{\rev{Residual ratios of estimated Pearson correlation coefficients to true Pearson correlation coefficients}}\label{tab:corrcompare}
		\end{table}
\rev{We acknowledge that the use of the two-step Bayesian inference does not provide the full picture due to the dependence between the estimated parameters and the reserve being a non-linear function in terms of these parameters. However, this calibration approach was selected due to a number of advantages as mentioned in Section \ref{Sec:Model}. These include overcoming the difficulties in dealing with Tweedie densities which are not in tractable form, and enhancing computational speed. From the analyses provided above, we can conclude that:}
\begin{arcitem}
\item \rev{The calibration method can capture the dependence structure well. }
\item \rev{The resulting reserve predictions show no apparent bias and they are in line with the chain ladder predictions.}
\end{arcitem}
\rev{Therefore, even though we may not get the full picture, the above results give us confidence that this would not have a material impact on the performance of the calibration. }
		
	\subsection{A comparison of performances of multivariate Tweedie models on unbalanced data}\label{Sec:sim2}
	A natural question arises regarding the performance of the multivariate Tweedie approach for unbalanced data compared to the original multivariate Tweedie approach introduced in \citet*{AvTaVuWo16}. To be able to assess their performances more accurately, this comparison is performed on {a simulated} data set whose underlying model is known. True common shock contributions are also known and these serve as the benchmark for the comparison. 
	
	To not put any particular framework at a disadvantage, the  synthetic data used for this illustration is simulated from a mixture of models. We deliberately select a (extreme) data set to which neither of the frameworks is properly adapted. In particular, two loss triangles of ten development periods and ten accident periods are generated such that the dependence is strong in the first four development periods, and not as strong in the last six periods. The common shock components are generated with column-specific mean parameters $\alpha_j = c_j\sqrt{\nu_j^{(1)}\nu_j^{(2)}}$ with $c_j=0.5$ for $1\leq j\leq4$, and $c_j=0.02$ for $5\leq j\leq 10$. The second business line is also simulated to be longer-tailed than the first. {Similar to the previous illustration, each observation in the two triangles is drawn from the multivariate Tweedie model for unbalanced data represented in Section \ref{Sec:Model}. These observations are assumed to have been standardised for accident year effect for simplicity.} The two loss triangles are presented in Table \ref{Tab:sim3} and \ref{Tab:sim4} in Appendix \ref{Sec:data2}. \par 
	
\begin{figure}[htb]
		\centering
		\includegraphics[scale=0.85]{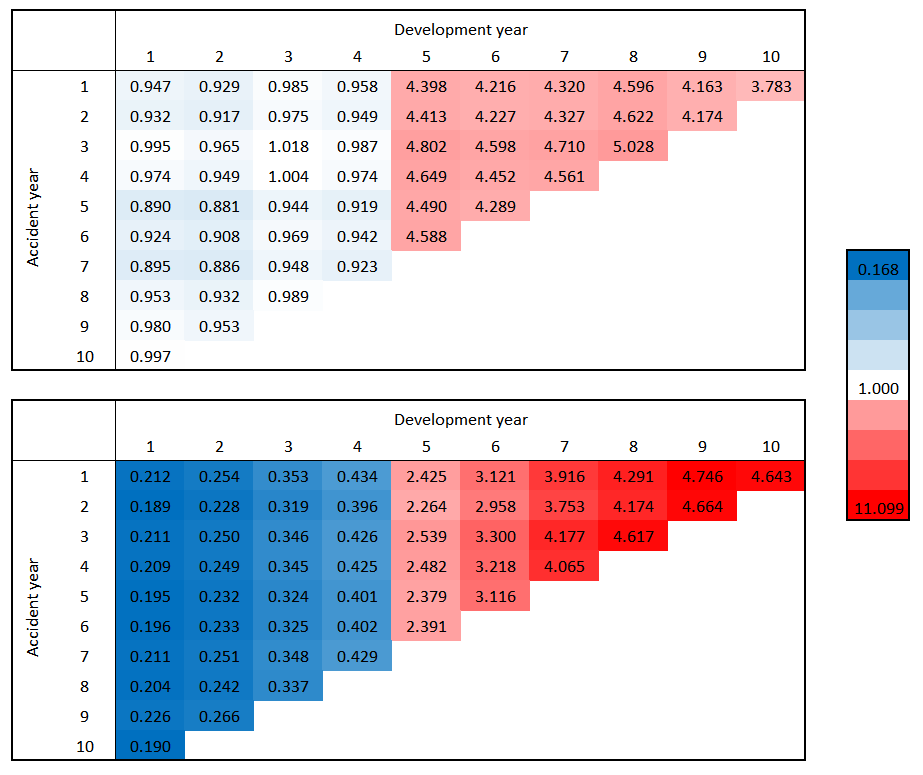}
		\caption{(Colour online) Heat maps of ratios of fitted common shock proportions to true proportions for triangle 1 (top: Tweedie framework modified for unbalanced data, bottom: original common shock Tweedie framework)}\label{fig:unbalancedcomparison1}
	\end{figure}

	 Heat maps of ratios of fitted common shock proportions to true proportions are given in Figure \ref{fig:unbalancedcomparison1} for triangle 1. Fitted values are calculated using {posterior median of parameters} and true values are calculated using true parameter values. The modified Tweedie model provides a very good fit for the first four development periods. The goodness of fit is  considerably less satisfactory in the later development periods when the true common shock proportion drops. The discrepancy is more significant for the first business line which has shorter tail development. The original common shock Tweedie model provides a poor goodness-of-fit overall, especially in early development periods. The proportions of common shock are underestimated in early development periods and overestimated in later periods. {Even though not reported here, similar results are also observed in heat maps of ratios of fitted common shock proportions to true proportions for triangle 2.}\par
	
	Overall the modified Tweedie framework does not fully eliminate the issues of unbalanced data across development periods, however there is a reduction. The fitting is quite good in early development periods, but is less satisfactory in later periods. The use of the geometric average of column factors across multiple triangles may contribute to this performance as the geometric average may not be close to some individual column factors if the development patterns are too different. However, it is worth emphasising that the example used has quite an extreme variation in common shock proportions across development periods, and one should not expect such radical variation in practice. In addition, the poor performance also arises from the discrepancy between the modified model with a constant scaling term $c$  and the true model generating the data (with column specific scaling term $c_j$). With this specification, it is not surprising that the earlier (large) development periods dominate the estimation of $c$. We do not expect good results because of model misspecification, but we can arrive at two main conclusions: the modified framework out-performs the original framework; and the common shock proportions are mis-estimated in the higher development periods, where amounts are small and do not contribute significantly to total liability. It is also worth noting that the poor estimation of common shock proportion does not affect mean forecasts, only dependency between the triangles, and then only where the magnitudes of the forecasts are small.

\section{Illustration with real data}\label{Sec:realdata}
The data used for illustration is a set of two triangles from the Bodily Injury line (1) and the Accident Benefit (excluding Disability Income) line (2) from a Canadian insurance company provided in \citet*{CoGeAb16}. These two triangles have also been used for illustrations in Sections \ref{Sec:Intro} and \ref{Sec:Challenges} and their details can be found therein.

\subsection{Preliminary analysis}
A preliminary analysis is performed to assess the suitability of this data set. This includes the assessment of the tails, as well as the dependence structure. 

\subsubsection{Analysis of the tails}
 From the plots of loss ratios provided earlier in Figure \ref{fig:unbalancedplot}, it can be observed that the Bodily Injury line has longer claims development than the Accident Benefits line. Tail lengths of the two business lines are  also assessed using age-to-age development factors
\begin{equation}
 {f_j^{(n)} = \dfrac{\sum\limits_{i=1}^{I-j}Y_{i,j+1}^{(n)}}{\sum\limits_{i=1}^{I-j}Y_{i,j}^{(n)}}.}
\end{equation}
Results are given in Table \ref{Tab:tail}. It can be observed that the development factors of the Bodily Injury dominate those of the Accident Benefits line for all development periods, except in the final year. However, this blip may be a false signal due to the truncation of data at the last development period and only one single observation is made in this final year. Hence the Bodily Injury line is convincingly longer-tailed than the Accident Benefits line.  
\begin{table}[h!]
	\centering
	\begin{tabular}{cccccccccc}
		\toprule
		\hline
		$j$  & 1 & 2& 3&4&5&6&7&8&9\\
		\hline
		$f_{j}^{(1)}$ & 8.1617	&1.8968&	1.4521&	1.2652&	1.1249&	1.0624&	1.0225&	1.0254&	1.0092
		\\
		$f_{j}^{(2)}$ &2.5844&	1.3584&	1.1708&	1.1140&	1.0481&	1.0305&	1.0137&	1.0057&	1.0118
		\\
		\hline
		\bottomrule
	\end{tabular}
	\caption{Claims development factors for each development period}\label{Tab:tail}
\end{table} 

\subsubsection{Explanatory dependence analysis}\label{Sec:realdep}
A heuristic dependence analysis is performed by fitting to each line a Tweedie GLM with a log-link and the chain ladder mean structure
\begin{equation}
 {a_i^{(n)} + b_j^{(n)}.}
\end{equation}
This is to remove fixed accident period and development period effects. Correlations between GLM Pearson residuals of the two lines are given in Table \ref{Tab:corcal}. The dependence between residuals is strong and significant after allowing for fixed accident period and development period effects.\par 

\begin{table}[H]
	\centering
	\begin{tabular}{ccc}
		\toprule
		\hline
		Pearson & Spearman & Kendall \\ 
		\hline
		0.3659 (0.0060) & 0.3480 (0.0096)& 0.2525 (0.0065) \\ 
		\hline
		\bottomrule
	\end{tabular}
	\caption{Correlation coefficients between cell-wise GLM residuals and their corresponding $p$-values}
	\label{Tab:corcal}
\end{table}

To examine whether this strong correlation comes from calendar year effects that can impact both lines simultaneously, we also perform another GLM analysis with an additional fixed calendar year effect in the mean structure
\begin{equation}
 {a_i^{(n)} + b_j^{(n)} + h_t^{(n)}.\label{eq:unbalancedmeanglm}}
\end{equation}
Correlations between GLM Pearson residuals of the two lines are then given in Table \ref{Tab:corcalca}. The correlation coefficients have been reduced, however, not very significantly. \par 

\begin{table}[H]
	\centering
	\begin{tabular}{ccc}
		\toprule
		\hline
		Pearson & Spearman & Kendall \\ 
		\hline
		0.3416 (0.0107) & 0.3250 (0.0159)& 0.2202 (0.0176) \\ 
		\hline
		\bottomrule
	\end{tabular}
	\caption{Correlation coefficients between cell-wise GLM residuals and their corresponding $p$-values after removing fixed calendar year effects}
	\label{Tab:corcalca}
\end{table}
Heat maps of {residual ratios} are given in Figure \ref{Fig:heatcal} in Appendix \ref{Sec:real}. {Residual ratios are defined as ratios of observed values to GLM fitted values with the mean structure specified in Equation \eqref{eq:unbalancedmeanglm}}. There are some common cell-wise patterns that are quite obvious from the heat maps, for example, low payments in development year 7 compensated by accelerated payments in years 8-9 in the first accident year, payment dips in accident year 4 and development year 2, similar development patterns in accident years from the preliminary analysis shows that this data set is suitable for illustration of the model. 

Results from the preliminary analysis shows that this data set is suitable to be used for illustration of the model. 

\subsection{{Estimation and goodness-of-fit assessment}}
Bayesian inference is used for estimation. The marginal fitting is first performed. 400,000 simulations are run and 300,000 simulations are discarded as the burn-in period. The sample chain is thinned by accepting every $5^{\text{th}}$ iteration to reduce the serial dependence between iterations. The multivariate fitting is then performed with 90,000 simulations and the first 30,000 are discarded as the burn-in period. The chain is then thinned by selecting every $3^{\text{th}}$ iteration. Summary statistics are then computed on these posterior samples. The results are given in Table \ref{Tab:marreal} and \ref{Tab:mulreal} {of} {Appendix \ref{Sec:real}.}\par 

\begin{figure}[H]
	\centering
	\includegraphics[scale=0.63]{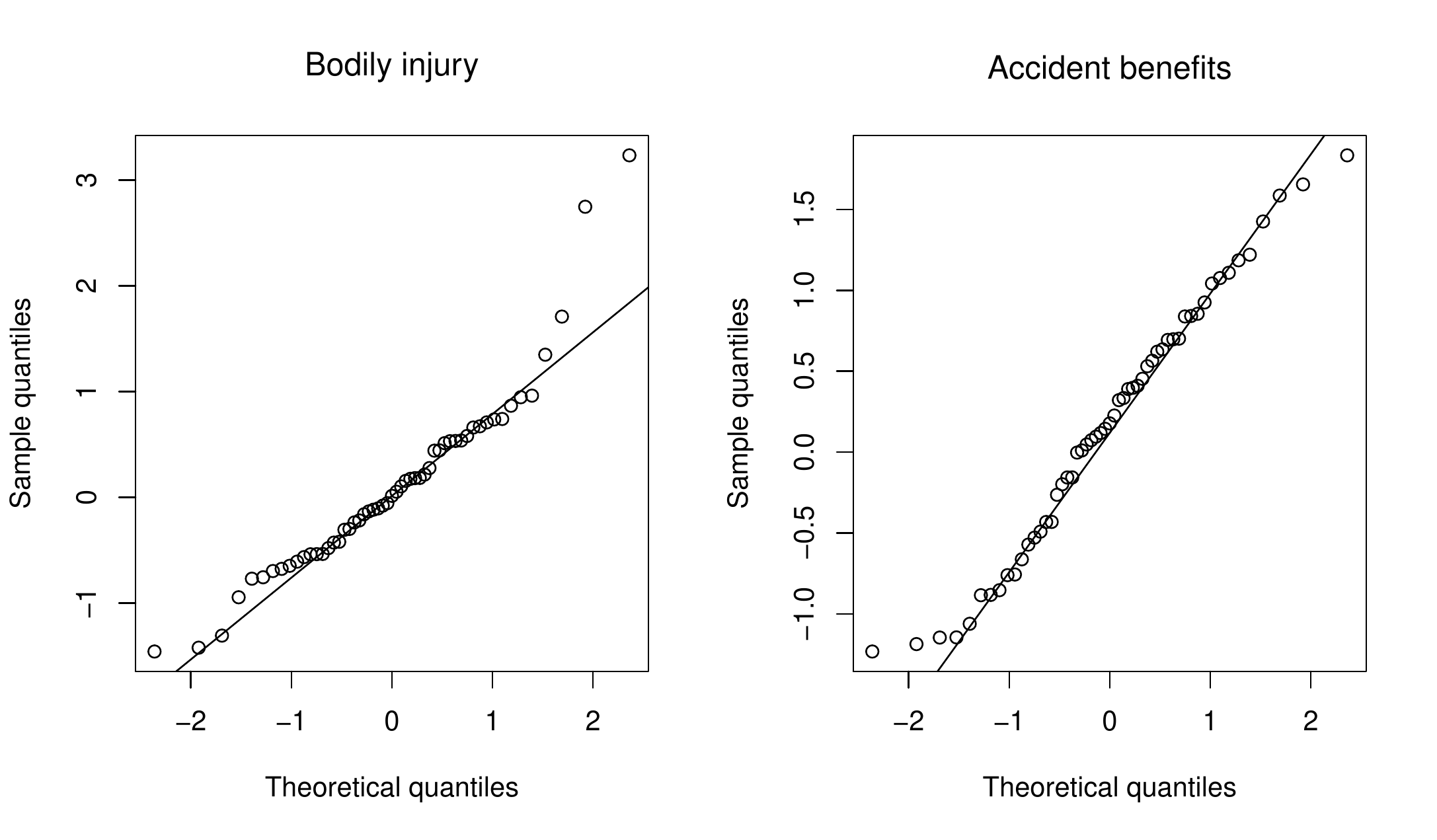}
	\caption{QQ plots of residuals from common shock Tweedie model ($p=1.829$)}\label{Fig:qqtwe}
\end{figure}

\begin{figure}[H]
	\centering
	\includegraphics[scale=0.63]{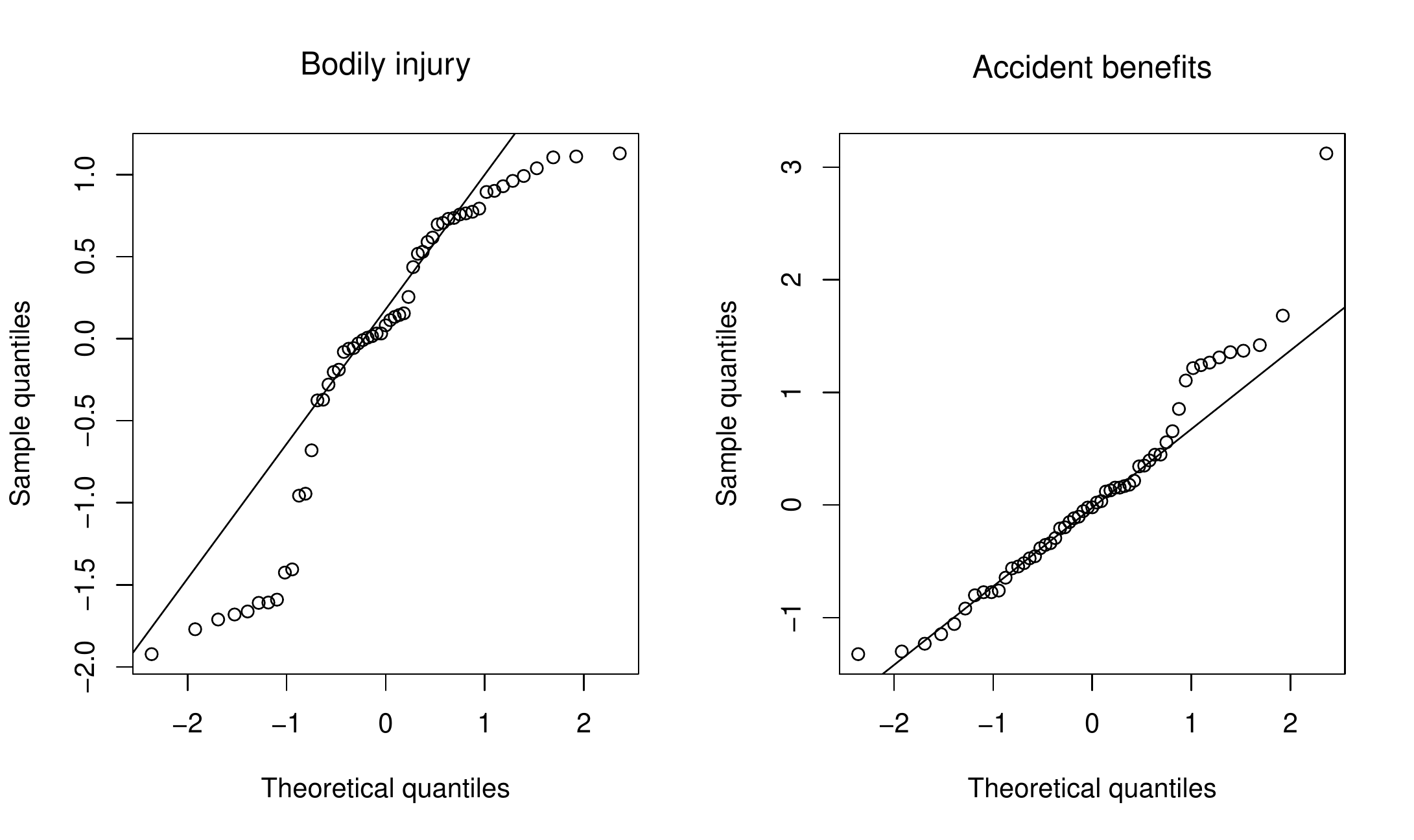}
	\caption{QQ plots of residuals from common shock normal model}\label{Fig:qqlog}
\end{figure}

Marginal and multivariate goodness-of-fits are assessed. Marginal goodness-of-fit is assessed using QQ plots of residuals in Figure \ref{Fig:qqtwe}. The plot shows that the fit is quite off in the right tail of the Bodily Injury line, and slightly off in both tails of the Accident Benefit line. The goodness of fit in other regions, however, is reasonable. This may be a result of the restriction of using the same power parameter $p$ for both lines. However, the multivariate Tweedie framework still provides marginal flexibility with flexible choices of $p$. For comparison, similar QQ plots are performed for a common shock normal model in Figure \ref{Fig:qqlog}. It can be observed that the Tweedie marginals provide a much better fit than the normal marginals (with power parameter  $p=0$). \par 


\begin{figure}[H]
	\centering
	\includegraphics[scale=0.65]{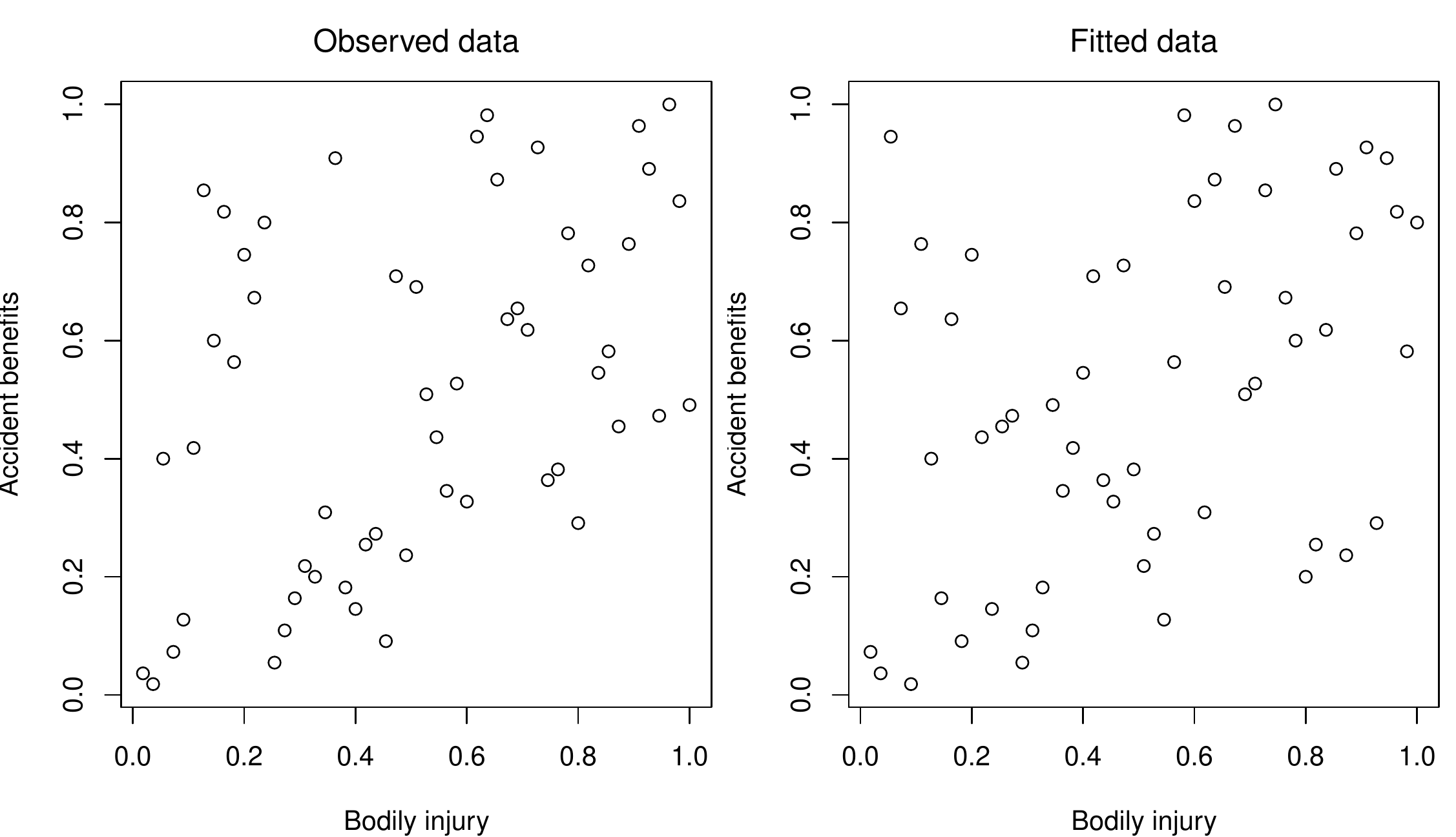}
	\caption{Plots of {empirical bivariate marginals} for observed values and back-fitted values}\label{Fig:cop}
\end{figure}

Multivariate goodness-of-fit is assessed by comparing {the empirical bivariate marginals of real data observations and of back fitted values. These are obtained using the empirical cumulative distribution functions of claim observations from each triangle.} Because of the use of a Bayesian inference, various sets of back fitted data can be generated. A path is randomly chosen for illustration. {Scatter plots of these empirical bivariate marginals} are presented in Figure \ref{Fig:cop}. It can be observed that the model can capture the general positive dependence structure in the data. 

{To look for any trace of dependence not captured by the model, we examine the residuals from model fitting. These residuals are obtained {as the differences between observations and fitted values, where the latter are calculated using posterior estimates}. The Pearson correlation coefficient of these residuals reduces to 0.1204 (p-value 0.3812). This is much weaker than the correlation coefficient of 0.3416 of GLM Pearson residuals in Section \ref{Sec:realdep} and is also insignificant. The insignificant  correlation indicates that our model has explained away most of the dependence in the data.}

\subsection{Common shock proportions}
Predictive distributions of outstanding claim observations in the lower triangles can be calculated using predictive Bayesian inference. Using parameter estimates, the contributions of common shock within each cell in the two triangles are calculated and given in Table \ref{Tab:shockprop1real} and \ref{Tab:shockprop2real}. It can be observed that there is only a very mild variation in the common shock proportions within and across triangles. {We can relate this result to the challenges coming from applying a common shock model to loss reserving data which has an unbalanced nature discussed in Section \ref{Sec:Challenges}. It shows that the proposed approach has provided a balance of common shock proportions across all loss cells within and across loss triangles.}

\begin{table}[htb]
	\centering
	\begin{tabular}{cc|cccccccccc}
		\toprule
		\hline
		& & \multicolumn{10}{c}{Development year}\\
		&	& 1 & 2 & 3 & 4 & 5 & 6 & 7 & 8 & 9 & 10 \\
		\hline 
		\multirow{10}{*}{\rotatebox[origin=c]{90}{Accident year}}&
		1 & 4.8\% & 4.2\% & 4.1\% & 4.0\% & 3.9\% & 3.9\% & 4.0\% & 4.1\% & 3.6\% & 4.2\% \\ 
	&	2 & 5.2\% & 4.6\% & 4.4\% & 4.3\% & 4.2\% & 4.2\% & 4.3\% & 4.4\% & 3.9\% & 4.5\% \\ 
	&	3 & 4.9\% & 4.3\% & 4.2\% & 4.0\% & 3.9\% & 4.0\% & 4.1\% & 4.2\% & 3.7\% & 4.2\% \\ 
	&	4 & 5.1\% & 4.5\% & 4.3\% & 4.2\% & 4.1\% & 4.2\% & 4.2\% & 4.3\% & 3.8\% & 4.4\% \\ 
	&	5 & 4.9\% & 4.3\% & 4.1\% & 4.0\% & 3.9\% & 4.0\% & 4.0\% & 4.1\% & 3.6\% & 4.2\% \\ 
	&	6 & 5.0\% & 4.4\% & 4.2\% & 4.1\% & 4.0\% & 4.1\% & 4.1\% & 4.2\% & 3.7\% & 4.3\% \\ 
	&	7 & 4.7\% & 4.1\% & 4.0\% & 3.8\% & 3.7\% & 3.8\% & 3.9\% & 4.0\% & 3.5\% & 4.0\% \\ 
	&	8 & 5.0\% & 4.3\% & 4.2\% & 4.1\% & 4.0\% & 4.1\% & 4.1\% & 4.2\% & 3.7\% & 4.3\% \\ 
	&	9 & 5.1\% & 4.5\% & 4.3\% & 4.2\% & 4.1\% & 4.2\% & 4.2\% & 4.3\% & 3.8\% & 4.4\% \\ 
	&	10 & 6.2\% & 5.4\% & 5.3\% & 5.1\% & 5.0\% & 5.1\% & 5.1\% & 5.2\% & 4.6\% & 5.3\% \\ 
		\hline
		\bottomrule
	\end{tabular}
	\caption{Proportions of common shock to the expected total observations calculated using parameter estimates - Bodily Injury}\label{Tab:shockprop1real}
\end{table}

\begin{table}[htb]
	\centering
	\begin{tabular}{cc|cccccccccc}
		\toprule
		\hline
		& & \multicolumn{10}{c}{Development year}\\
		&	& 1 & 2 & 3 & 4 & 5 & 6 & 7 & 8 & 9 & 10 \\
		\hline 
		\multirow{10}{*}{\rotatebox[origin=c]{90}{Accident year}}&
		1 & 4.4\% & 5.0\% & 5.1\% & 5.3\% & 5.4\% & 5.4\% & 5.3\% & 5.2\% & 5.9\% & 5.1\% \\ 
	&	2 & 4.6\% & 5.2\% & 5.3\% & 5.5\% & 5.7\% & 5.6\% & 5.5\% & 5.4\% & 6.1\% & 5.3\% \\ 
	&	3 & 4.4\% & 5.0\% & 5.1\% & 5.3\% & 5.4\% & 5.3\% & 5.3\% & 5.1\% & 5.9\% & 5.1\% \\ 
	&	4 & 4.2\% & 4.7\% & 4.9\% & 5.1\% & 5.2\% & 5.1\% & 5.0\% & 4.9\% & 5.6\% & 4.8\% \\ 
	&	5 & 4.4\% & 5.0\% & 5.1\% & 5.3\% & 5.4\% & 5.4\% & 5.3\% & 5.2\% & 5.9\% & 5.1\% \\ 
	&	6 & 4.1\% & 4.7\% & 4.8\% & 5.0\% & 5.1\% & 5.1\% & 5.0\% & 4.9\% & 5.5\% & 4.8\% \\ 
	&	7 & 4.1\% & 4.7\% & 4.8\% & 5.0\% & 5.1\% & 5.0\% & 4.9\% & 4.8\% & 5.5\% & 4.7\% \\ 
	&	8 & 4.2\% & 4.8\% & 5.0\% & 5.1\% & 5.3\% & 5.2\% & 5.1\% & 5.0\% & 5.7\% & 4.9\% \\ 
	&	9 & 4.2\% & 4.8\% & 5.0\% & 5.1\% & 5.3\% & 5.2\% & 5.1\% & 5.0\% & 5.7\% & 4.9\% \\ 
	&	10 & 4.0\% & 4.6\% & 4.7\% & 4.9\% & 5.0\% & 4.9\% & 4.8\% & 4.7\% & 5.4\% & 4.6\% \\ 
		\hline
		\bottomrule
	\end{tabular}
	\caption{Proportions of common shock to the expected total observations calculated using parameter estimates - Accident Benefits}\label{Tab:shockprop2real}
\end{table}

\subsection{Outstanding claims forecast}
\begin{figure}[H]
	\centering
	\includegraphics[scale=0.68]{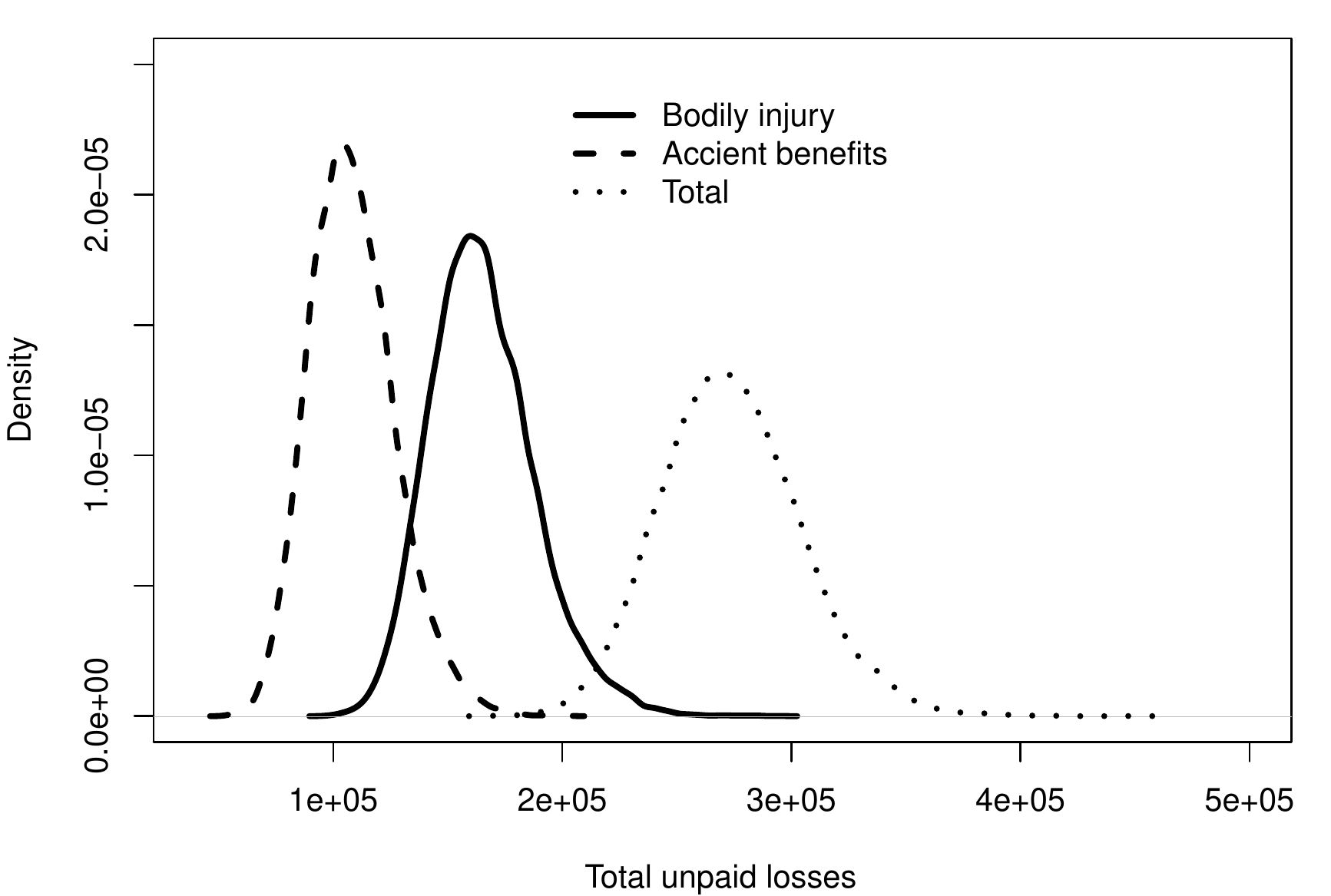}
	\caption{Kernel densities of predictive distributions of total outstanding claims in each line of business and in the aggregate portfolio}\label{Fig:kernel}
\end{figure}

 To obtain the distributions of the outstanding claims, posterior samples of parameters from the Bayesian inference are used to project claims in lower triangles. This projection utilises the specification in Equations \eqref{zmarginalreprod}, \eqref{Eq:shock} and \eqref{Eq:modelstructure}. This gives a set of samples of future claims in the lower triangles. Using this set, summary statistics of the total outstanding claims distributions are given in Table \ref{Tab:sumstat} and kernel densities of outstanding claims are given in Figure \ref{Fig:kernel}. Summary statistics provided include the posterior mean, standard deviation, $\text{VaR}_{75\%}$ and $\text{VaR}_{95\%}$ of the distribution of total outstanding claims for each line, as well as for both lines. \par 

\begin{figure}[H]
	\centering
	\includegraphics[scale=0.6]{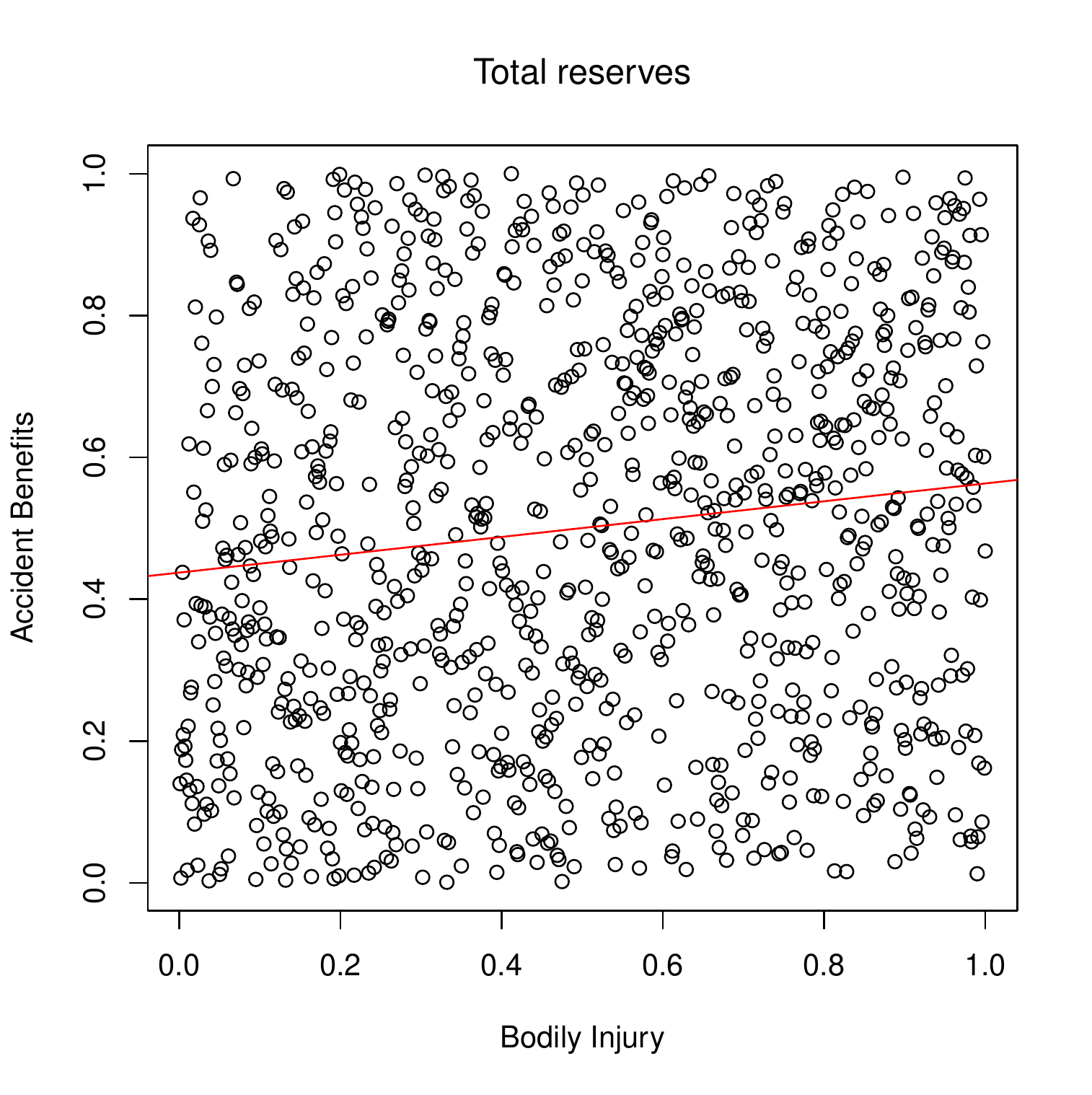}
	\caption{Plot of empirical bivariate marginals of total reserves (using 1,000 posterior samples)}\label{fig:reserves_empirical}
\end{figure}

\begin{table}[H]
	\centering
	\begin{tabular}{lrrr}
		\toprule
		\hline  
		& \multicolumn{1}{c}{Bodily Injury} & \multicolumn{1}{c}{Accident Benefits} & \multicolumn{1}{c}{Both lines} \\ 
		\hline
		Mean & 165,185.92 & 108,465.81  & 273,651.73  \\ 
		SD & 22,720.88  & 18,554.65  & 30,538.83  \\ 
		$\text{VaR}_{75\%}$ & 179,057.18 & 120,100.43 & 293,061.56  \\ 
		$\text{VaR}_{95\%}$ & 205,752.20  & 141,426.24 & 326,177.22  \\ 
		\hline
		\bottomrule
	\end{tabular}
	\caption{Summary statistics of outstanding claims distributions}\label{Tab:sumstat}
\end{table}
{The empirical bivariate marginals of total reserves are shown in Figure \ref{fig:reserves_empirical}. For illustration purpose, we show the scatter plot of total reserves from 1,000 posterior samples. The plot shows a mild positive dependence structure in the total outstanding claims across two lines. This is accompanied by a Pearson correlation of 0.0855 (p-value $<$2.2e-16). There is no clear evidence of a concentration in the tail regions of the dependence. There can be diversification across claims within a single loss triangle. Hence the dependence on the aggregate reserves from each line is mild and tail dependence may not be apparent.}
The two business lines do not have a comonotonic dependence structure, and this allows the insurer to gain some diversification benefits when they set their risk margins. Using the specification of a risk margin under APRA's Prudential Standards GPS 340, we have the following definition of Risk margin and Diversification Benefit (DB) 
\begin{align}
&\text{Risk margin}_{\chi\%}[Y] = \max\left\lbrace \text{VaR}_{\chi\%}[Y] - \text{E}[Y];\,\dfrac{1}{2}\text{SD}[Y]\right\rbrace,\\
& \text{DB} = \dfrac{\left(\text{Risk margin}_{\chi\%}[Y_1]+\text{Risk margin}_{\chi\%}[Y_2]\right) -  \text{Risk margin}_{\chi\%}[Y_1+Y_2]}{\text{Risk margin}_{\chi\%}[Y_1]+\text{Risk margin}_{\chi\%}[Y_2]} \times 100\%.
\end{align}
Risk Margin$_{75\%}$ and Risk Margin$_{95\%}$, as well as associated diversification benefits are provided in Table \ref{Tab:riskmargin}. It can then be observed that diversification benefits can be gained as a result of allowing for (non-comonotonic) dependence across business lines. 

\begin{table}[H]
	\centering
	\begin{tabular}{lrrrr}
		\toprule
		\hline
		& \multicolumn{1}{c}{Bodily Injury} & \multicolumn{1}{c}{Accident Benefits} & \multicolumn{1}{c}{Both lines} & \multicolumn{1}{c}{DB} \\
		\hline
		Risk margin$_{75\%}$  & 13,871.26 & 11,634.61 &19,409.83 & 23.9\%\\ 
		Risk margin$_{95\%}$ &  40,566.28 & 32,960.43 &52,525.49 & 28.6\%\\ 
		\hline
		\bottomrule
	\end{tabular}
	\caption{Risk margin and diversification benefits statistics}\label{Tab:riskmargin}
\end{table}

\section{Conclusion}\label{Sec:Conclusion}
 Common shock approaches can provide many benefits in the modelling of outstanding claims. However, they often require very careful parametrisation. This arises from the unbalanced nature of loss reserving data. It is often desirable to use scaling factors to adjust the common shock effects so that they can contribute proportionately to the total observations over the entire range of the triangles. However, an excessive use of scaling factors can result in over-parametrisation. In some cases, such as the common shock Tweedie framework developed in \citet*{AvTaVuWo16}, it is also desirable to select scaling factors such that distributional tractability is preserved. These requirements place conflicting constraints on the specification of scaling factors in common shock models. 

 In this paper, we propose an approach which compromises the various constraints mentioned above. This approach involves using careful and parsimonious parametrisation to develop a common shock Tweedie framework modified for unbalanced data. Additional modifications for negative claims are also undertaken under this framework. Illustrations using simulated and with real data are presented. These illustrations show that while the proposed approach cannot fully eliminate the issue of unbalanced common shock proportions, the improvement over the original framework in \citet*{AvTaVuWo16} is quite substantial.\par 

 We examined a common shock Tweedie approach for cell-wise dependence in this paper. Future research could consider applications on other structures of dependence (such as calendar period dependence). This paper raises some potential issues of common shock models when they are applied to reserving data that has an unbalanced nature. These issues, however, might appear whenever common shock models are applied to heterogeneous data. These can include mortality data for different group ages, or capital modelling for different types of risks. The proposed solution could be extended to solve similar problems in other contexts. While this solution can reduce the problems of unbalanced data quite substantially, a complete balance in common shock proportions cannot be achieved. Future research could consider a better solution to this problem. Other multivariate models with explicit dependence structures such as mixture models could also be considered as they might be more applicable to unbalanced data.
 
\section*{Acknowledgements}
Results in this paper were presented at The Australasian Actuarial Education and Research Symposium in 2017 and the 22nd International Congress on Insurance: Mathematics and Economics in 2018. The authors are grateful for constructive comments received from colleagues who attended these conferences. {The authors are also thankful to the two anonymous reviewers for their constructive comments that helped significantly improve the paper.}

This research was  supported under Australian Research Council's Linkage (LP130100723, with funding partners Allianz Australia Insurance Ltd, Insurance Australia Group Ltd, and Suncorp Metway Ltd) and Discovery (DP200101859) Projects funding schemes. Furthermore, Phuong Anh Vu acknowledges financial support from a University International Postgraduate Award/University Postgraduate Award and supplementary scholarships provided by the UNSW Business School. The views expressed herein are those of the authors and are not necessarily those of the supporting organisations.

\bibliographystyle{elsarticle-harv}
\bibliography{Reserving}

\newpage
\appendix
\gdef\thesection{\Alph{section}}
\makeatletter
\renewcommand\@seccntformat[1]{Appendix \csname the#1\endcsname.\hspace{0.5em}}
\makeatother
	\setcounter{table}{0}
\renewcommand{\thetable}{A.\arabic{table}}
\section{Simulated data set 1}\label{Sec:data1}

\begin{table}[H]
	\centering
		\begin{tabular}{cc|cccccccccc}
		\toprule
		\hline
		& & \multicolumn{10}{c}{Development year}\\
		&	& 1 & 2 & 3 & 4 & 5 & 6 & 7 & 8 & 9 & 10 \\
		\hline 
		\multirow{10}{*}{\rotatebox[origin=c]{90}{Accident year}}&
		1 & 85.57 & 43.18 & 20.58 & 13.40 & 4.40 & 2.34 & 1.86 & 0.55 & 0.28 & 0.15 \\ 
	&	2 & 78.22 & 28.65 & 12.74 & 5.08 & 6.97 & 2.82 & 1.50 & 0.07 & -0.01 &  \\ 
	&	3 & 85.90 & 36.58 & 22.21 & 14.29 & 2.23 & 3.31 & 0.82 & 1.86 &  &  \\ 
	&	4 & 67.86 & 36.94 & 16.01 & 11.23 & 5.54 & 4.68 & 1.40 &  &  &  \\ 
	&	5 & 83.45 & 33.30 & 21.24 & 10.80 & 4.32 & 3.04 &  &  &  &  \\ 
	&	6 & 63.85 & 39.38 & 24.71 & 2.84 & 7.77 &  &  &  &  &  \\ 
	&	7 & 78.80 & 31.17 & 16.96 & 8.27 &  &  &  &  &  &  \\ 
	&	8 & 90.32 & 36.19 & 13.56 &  &  &  &  &  &  &  \\ 
	&	9 & 97.94 & 35.43 &  &  &  &  &  &  &  &  \\ 
	&	10 & 58.14 &  &  &  &  &  &  &  &  &  \\ 
		\hline
		\bottomrule
	\end{tabular}
	\caption{Simulated triangle 1 (data set 1)}\label{Tab:sim1}
\end{table}

\begin{table}[H]
	\centering
	\begin{tabular}{cc|cccccccccc}
	\toprule
	\hline
	& & \multicolumn{10}{c}{Development year}\\
	&	& 1 & 2 & 3 & 4 & 5 & 6 & 7 & 8 & 9 & 10 \\
	\hline 
	\multirow{10}{*}{\rotatebox[origin=c]{90}{Accident year}}&
		1 & 24.12 & 38.93 & 45.70 & 43.19 & 16.04 & 8.70 & 4.78 & 1.83 & 1.45 & 1.66 \\ 
	&	2 & 21.04 & 40.05 & 35.83 & 19.93 & 15.27 & 11.21 & 6.84 & 2.81 & 1.12 &  \\ 
	&	3 & 23.98 & 38.59 & 40.73 & 47.22 & 22.01 & 10.36 & 3.49 & 3.53 &  &  \\ 
	&	4 & 26.34 & 42.48 & 57.27 & 29.72 & 24.03 & 12.11 & 1.86 &  &  &  \\ 
	&	5 & 29.46 & 33.18 & 44.63 & 39.51 & 25.97 & 11.60 &  &  &  &  \\ 
	&	6 & 23.67 & 48.70 & 49.66 & 20.12 & 21.34 &  &  &  &  &  \\ 
	&	7 & 29.10 & 36.51 & 50.52 & 43.98 &  &  &  &  &  &  \\ 
	&	8 & 30.58 & 53.40 & 49.21 &  &  &  &  &  &  &  \\ 
	&	9 & 31.16 & 50.48 &  &  &  &  &  &  &  &  \\ 
	&	10 & 31.04 &  &  &  &  &  &  &  &  &  \\ 
		\hline
		\bottomrule
	\end{tabular}
	\caption{Simulated triangle 2 (data set 1)}\label{Tab:sim2}
\end{table}
\begin{landscape}

\begin{table}[h!]
	\centering
	\begin{tabular}{lcccc|lcccc}
		\toprule
		\hline
		& True value & Median & SD & 90\% CI & & True value & Median & SD & 90\% CI \\ 
		\hline
		$\eta_2^{(1)}$ &1.0300 & 0.8950 & 0.0756 & (0.8250; 1.0610) & $\eta_2^{(2)}$ &  1.1900 & 1.0620 & 0.1285 & (0.8850; 1.3050  ) \\ 
		$\eta_3^{(1)}$ & 1.1900 & 1.2820 & 0.1523 & (1.0530; 1.5560) & $\eta_3^{(2)}$& 1.1700 & 1.2300 & 0.1693 & (0.9950; 1.5440) \\ 
		$\eta_4^{(1)}$ & 1.1200 & 1.0250 & 0.1190 & (0.8590; 1.2460 ) & $\eta_4^{(2)}$&1.1500 & 1.1020 & 0.1507 & (0.8860; 1.3810  ) \\ 
		$\eta_5^{(1)}$ &  1.1500 & 1.0600 & 0.1200 & (0.8790; 1.2700) & $\eta_5^{(2)}$&1.1500 & 1.3190 & 0.1635 & (1.0710; 1.6040) \\ 
		$\eta_6^{(1)}$ & 1.1600 & 1.1740 & 0.1468 & (0.9650; 1.4460 ) & $\eta_6^{(2)}$& 1.2000 & 1.1390 & 0.1421 & (0.9340; 1.3970 ) \\ 
		$\eta_7^{(1)}$ & 1.1200 & 1.0010 & 0.1159 & (0.8530; 1.2270) & $\eta_7^{(2)}$&  1.4000 & 1.4480 & 0.1912 & (1.1690; 1.7950) \\ 
		$\eta_8^{(1)}$ & 1.1400 & 1.0380 & 0.1219 & (0.8690; 1.2700) & $\eta_8^{(2)}$& 1.4500 & 1.5000 & 0.2095 & (1.1890; 1.8750) \\ 
		$\eta_9^{(1)}$ & 1.2100 & 1.1310 & 0.1343 & (0.9350; 1.3760 ) & $\eta_9^{(2)}$& 1.5600 & 1.4820 & 0.1908 & (1.2110; 1.8320 ) \\ 
		$\eta_{10}^{(1)}$ & 1.1900 & 1.0680 & 0.1879 & (0.8530; 1.4470) & $\eta_{10}^{(2)}$&1.6600 & 1.9840 & 0.5113 & (1.3110; 2.9670  ) \\ 
		\hline
		$\nu_1^{(1)}$ & 60.0000 & 56.8610 & 4.2527 & (49.8400; 63.7100) & $\nu_1^{(2)}$ & 10.0000 & 10.2800 & 1.1138 & (8.5280; 12.1490) \\ 
		$\nu_2^{(1)}$ & 20.0000 & 21.7210 & 1.9812 & (18.6780; 25.1880 ) & $\nu_2^{(2)}$ &  20.0000 & 22.1790 & 2.0222 & (18.9970; 25.6420) \\ 
		$\nu_3^{(1)}$ & 10.0000 & 10.0160 & 1.3134 & (8.1100; 12.4490 ) & $\nu_3^{(2)}$& 25.0000 & 27.2130 & 2.5612 & (23.1110; 31.5170) \\ 
		$\nu_4^{(1)}$ & 5.0000 & 4.6610 & 0.7082 & (3.5870; 5.8940) & $\nu_4^{(2)}$& 20.0000 & 21.2330 & 2.1120 & (17.9160; 24.8910) \\ 
		$\nu_5^{(1)}$ &  2.5000 & 2.2920 & 0.2783 & (1.8700; 2.7830) & $\nu_5^{(2)}$&  15.0000 & 13.9600 & 1.6821 & (11.4140; 16.8640) \\ 
		$\nu_6^{(1)}$ & 1.2500 & 1.5100 & 0.2095 & (1.2120; 1.892) & $\nu_6^{(2)}$& 8.0000 & 6.5560 & 0.9920 &(5.0850; 8.33707) \\ 
		$\nu_7^{(1)}$ &  0.6000 & 0.7290 & 0.1453 & (0.5320; 1.0040 ) & $\nu_7^{(2)}$&  3.0000 & 2.7070 & 0.4708 & (2.0370; 3.5830) \\ 
		$\nu_8^{(1)}$ & 0.3000 & 0.2620 & 0.0449 & (0.2010; 0.3460) & $\nu_8^{(2)}$& 2.0000 & 1.9430 & 0.3335 & (1.4350; 2.5340) \\ 
		$\nu_9^{(1)}$ & 0.1500 & 0.1430 & 0.0272 & (0.1040; 0.1930) & $\nu_9^{(2)}$&1.0000 & 0.7700 & 0.1122 & (0.6180; 1.0100 ) \\ 
		$\nu_{10}^{(1)}$ & 0.1500 & 0.2040 & 0.0445 & (0.1440; 0.2860 ) & $\nu_{10}^{(2)}$&  1.0000 & 0.9480 & 0.1997 & (0.6580; 1.3140) \\ 
		\hline
		$\gamma^{(1)}$ &  0.5000 & 0.5530 & 0.0649 & (0.4560; 0.6680) & $\gamma^{(2)}$ & 0.7000 & 0.6220 & 0.0936 & (0.4880; 0.7930) \\ 
		$p$  & 1.3000 & 1.3640 & 0.0507 & (1.2850; 1.4510) & $\xi^{(1)}$ & 0.0100 & 0.0120 & 0.0013 & (0.0100; 0.0140)  \\ 
		$\delta$ & 1.0118 & 1.0420 & 0.1293 & (0.8470; 1.2710) & &&&&  \\ 
		$c$ & 0.500 &  0.3620 & 0.6942 & (0.0600; 2.2370) & 	$\beta$ & 0.600 & 0.5030 & 0.4603 & (0.1610; 1.6020)\\
		\hline
		\bottomrule
	\end{tabular}
	\caption{Posterior statistics of parameters (data set 1)}\label{Tab:marginal1}
\end{table}
\end{landscape}
	\setcounter{figure}{0}
\renewcommand{\thefigure}{A.\arabic{figure}}

		\begin{figure}[H]
		\centering
		\includegraphics[scale=0.8]{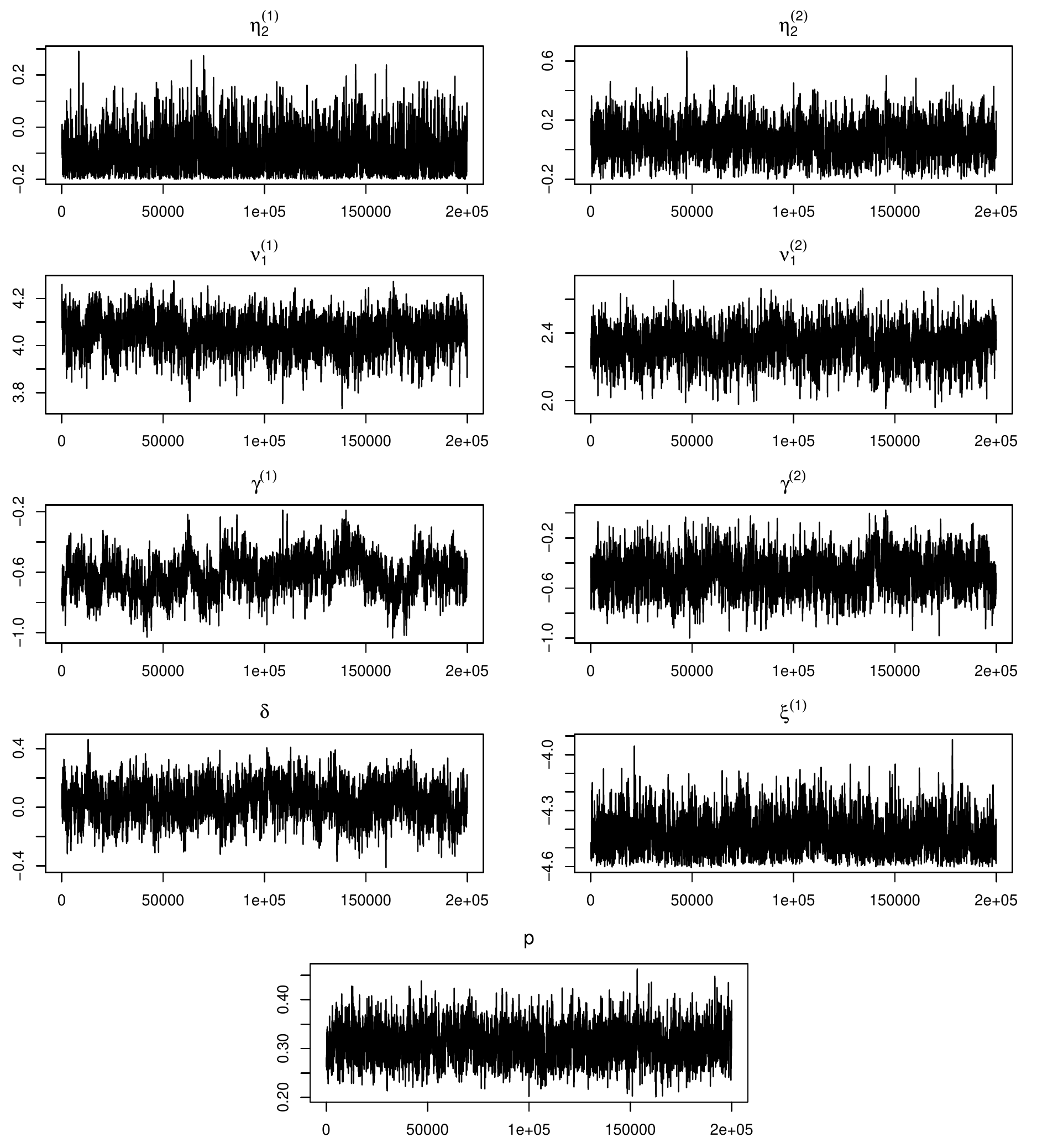}
		\caption{MCMC sample paths of some parameters}\label{Fig:mcmc1}
	\end{figure}
	\setcounter{table}{0}
\renewcommand{\thetable}{B.\arabic{table}}
\section{Simulated data set 2}\label{Sec:data2}
\begin{table}[H]
	\centering
	\begin{tabular}{cc|cccccccccc}
		\toprule
		\hline
		& & \multicolumn{10}{c}{Development year}\\
		&	& 1 & 2 & 3 & 4 & 5 & 6 & 7 & 8 & 9 & 10 \\
		\hline 
		\multirow{10}{*}{\rotatebox[origin=c]{90}{Accident year}}&
		1 & 47.16 & 36.33 & 18.58 & 10.63 & 3.93 & 0.38 & 0.45 & 0.00 & 0.00 & 0.13 \\ 
	&	2 & 102.30 & 49.12 & 18.47 & 17.05 & 3.30 & 1.70 & 1.77 & 1.04 & 0.00 &  \\ 
	&	3 & 101.87 & 56.91 & 14.75 & 24.29 & 1.46 & 1.15 & 0.83 & 0.17 &  &  \\ 
	&	4 & 97.09 & 35.96 & 27.80 & 10.86 & 3.93 & 3.71 & 0.43 &  &  &  \\ 
	&	5 & 107.07 & 34.34 & 20.49 & 19.73 & 6.65 & 1.70 &  &  &  &  \\ 
	&	6 & 107.10 & 66.55 & 27.03 & 17.09 & 2.38 &  &  &  &  &  \\ 
	&	7 & 123.60 & 37.41 & 32.77 & 15.53 &  &  &  &  &  &  \\ 
	&	8 & 107.03 & 50.50 & 18.30 &  &  &  &  &  &  &  \\ 
	&	9 & 105.93 & 42.02 &  &  &  &  &  &  &  &  \\ 
	&	10 & 109.09 &  &  &  &  &  &  &  &  &  \\ 
		\hline
		\bottomrule
	\end{tabular}
	\caption{Simulated triangle 1 (data set 2)}\label{Tab:sim3}
\end{table}

\begin{table}[H]
	\centering
\begin{tabular}{cc|cccccccccc}
	\toprule
	\hline
	& & \multicolumn{10}{c}{Development year}\\
	&	& 1 & 2 & 3 & 4 & 5 & 6 & 7 & 8 & 9 & 10 \\
	\hline 
	\multirow{10}{*}{\rotatebox[origin=c]{90}{Accident year}}&
		1 & 19.61 & 45.29 & 44.23 & 28.82 & 24.48 & 3.15 & 3.23 & 3.30 & 1.94 & 0.73 \\ 
	&	2 & 33.52 & 41.13 & 39.33 & 41.78 & 22.46 & 8.69 & 2.09 & 4.85 & 1.88 &  \\ 
	&	3 & 24.39 & 43.40 & 34.06 & 59.94 & 22.00 & 13.90 & 5.54 & 1.62 &  &  \\ 
	&	4 & 27.78 & 37.03 & 41.41 & 31.12 & 31.73 & 5.92 & 7.69 &  &  &  \\ 
	&	5 & 24.46 & 41.96 & 36.55 & 23.42 & 20.88 & 9.61 &  &  &  &  \\ 
	&	6 & 26.36 & 38.68 & 58.52 & 36.25 & 27.15 &  &  &  &  &  \\ 
	&	7 & 30.05 & 36.18 & 52.14 & 41.98 &  &  &  &  &  &  \\ 
	&	8 & 30.32 & 53.54 & 52.87 &  &  &  &  &  &  &  \\ 
	&	9 & 42.37 & 42.25 &  &  &  &  &  &  &  &  \\ 
	&	10 & 46.49 &  &  &  &  &  &  &  &  &  \\ 
			\hline
		\bottomrule
	\end{tabular}
	\caption{Simulated triangle 2 (data set 2)}\label{Tab:sim4}
\end{table}
	\setcounter{table}{0}
\renewcommand{\thetable}{C.\arabic{table}}
\section{Real data set}\label{Sec:real}
This data set is drawn from \citet*{CoGeAb16}.
	\begin{table}[H]
		\centering
	\begin{tabular}{cc|c|cccccccccc}
		\toprule
		\hline
	 &	&\multirow{2}{*}{Premium}& \multicolumn{10}{c}{Development year}\\
	&	&	& 1 & 2 & 3 & 4 & 5 & 6 & 7 & 8 & 9 & 10 \\
		\hline 
		\multirow{10}{*}{\rotatebox[origin=c]{90}{Accident year}}& 
			1 & 85,421 & 3,488 & 14,559 & 27,249 & 37,979 & 49,561 & 55,957 & 58,406 & 60,862 & 63,280 & 63,864 \\ 
	&		2 & 98,579 & 1,169 & 12,781 & 20,550 & 31,547 & 42,808 & 47,385 & 50,251 & 50,978 & 51,272 &  \\ 
	&		3 & 103,062 & 1,478 & 10,788 & 25,499 & 34,279 & 43,057 & 49,360 & 52,329 & 52,544 &  &  \\ 
	&		4 & 108,412 & 1,186 & 11,852 & 22,913 & 32,537 & 41,824 & 48,005 & 52,542 &  &  &  \\ 
	&		5 & 111,176 & 1,737 & 13,881 & 25,521 & 38,037 & 43,684 & 47,755 &  &  &  &  \\ 
	&		6 & 112,050 & 1,571 & 12,153 & 27,329 & 41,832 & 51,779 &  &  &  &  &  \\ 
	&		7 & 112,577 & 1,199 & 17,077 & 29,876 & 44,149 &  &  &  &  &  &  \\ 
	&		8 & 113,707 & 1,263 & 16,073 & 28,249 &  &  &  &  &  &  &  \\ 
	&		9 & 126,442 &  986 & 10,003 &  &  &  &  &  &  &  &  \\ 
	&		10 & 130,484 &  683 &  &  &  &  &  &  &  &  &  \\ 
			\hline
			\bottomrule
		\end{tabular}
		\caption{Bodily Injury line (cumulative claims)}\label{Tab:real1}
	\end{table}
	
	\begin{table}[H]
		\centering
	\begin{tabular}{cc|c|cccccccccc}
		\toprule
		\hline
		&	&\multirow{2}{*}{Premium}& \multicolumn{10}{c}{Development year}\\
		&	&	& 1 & 2 & 3 & 4 & 5 & 6 & 7 & 8 & 9 & 10 \\
		\hline 
		\multirow{10}{*}{\rotatebox[origin=c]{90}{Accident year}}& 
			1 & 116,491 & 13,714 & 24,996 & 31,253 & 38,352 & 44,185 & 46,258 & 47,019 & 47,894 & 48,334 & 48,902 \\ 
		&	2 & 111,467 & 6883 & 16,525 & 24,796 & 29,263 & 32,619 & 33,383 & 34,815 & 35,569 & 35,612 &  \\ 
		&	3 & 107,241 & 7933 & 22,067 & 32,801 & 38,028 & 44,274 & 44,948 & 46,507 & 46,665 &  &  \\ 
		&	4 & 105,687 & 7052 & 18,166 & 25,589 & 31,976 & 36,092 & 38,720 & 39,914 &  &  &  \\ 
		&	5 & 105,923 & 10,463 & 23,982 & 31,621 & 36,039 & 38,070 & 41,260 &  &  &  &  \\ 
		&	6 & 111,487 & 9697 & 28,878 & 41,678 & 47,135 & 50,788 &  &  &  &  &  \\ 
		&	7 & 113,268 & 11,387 & 37,333 & 48,452 & 55,757 &  &  &  &  &  &  \\ 
		&	8 & 121,606 & 12,150 & 32,250 & 40,677 &  &  &  &  &  &  &  \\ 
		&	9 & 110,610 & 5348 & 14,357 &  &  &  &  &  &  &  &  \\ 
		&	10 & 104,304 & 4,612 &  &  &  &  &  &  &  &  &  \\ 
			\hline
			\bottomrule
		\end{tabular}
		\caption{Accident Benefits (cumulative claims)}\label{Tab:real2}
	\end{table}
\renewcommand{\thefigure}{C.\arabic{figure}}
\begin{figure}[H]
	\begin{center}
		\includegraphics[scale=0.85]{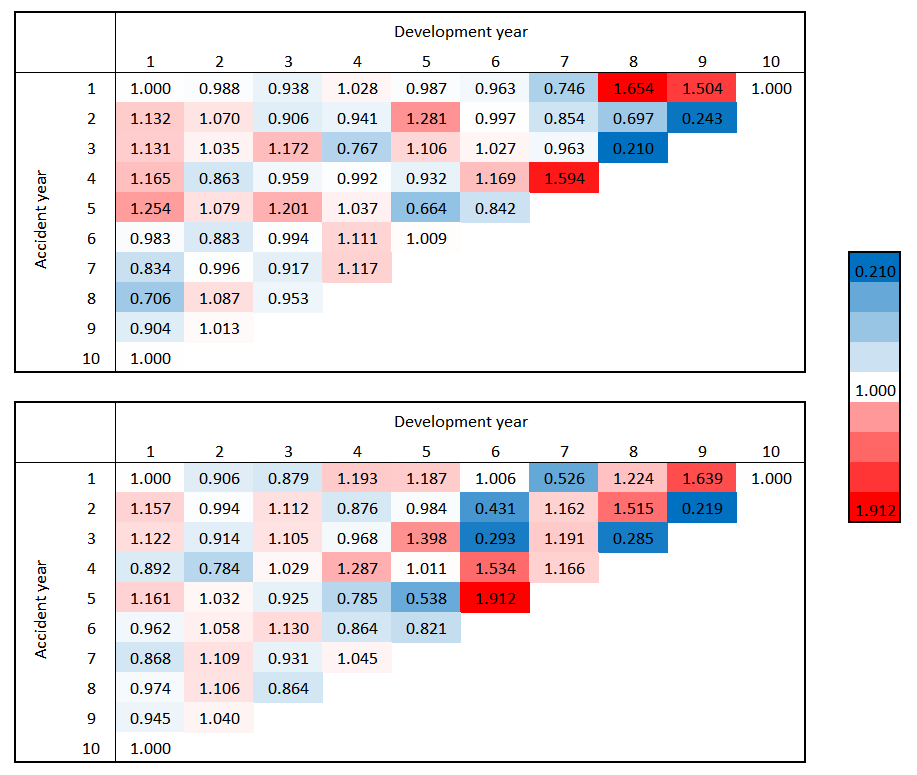}
	\end{center}
	\caption{(Colour online) Heat maps of ratios of observed values to GLM fitted values (top: Bodily Injury line, bottom: Accident Benefits)}\label{Fig:heatcal}
\end{figure}
\begin{table}[h!]
	\centering
	\begin{tabular}{lccc|lccc}
		\toprule
		\hline
		& Median & SD & 90\% CI &  & Median & SD & 90\% CI \\ 
		\hline
		$\eta_2^{(1)}$ & 0.6390 & 0.0875 & (0.5120; 0.7970 ) &$\eta_2^{(2)}$& 0.7770 & 0.1119 & (0.6230; 0.9830) \\ 
		$\eta_3^{(1)}$ & 0.9010 & 0.1026 & (0.7510; 1.0860) &$\eta_3^{(2)}$& 1.0050 & 0.1508 & (0.7860; 1.2770) \\ 
		$\eta_4^{(1)}$ &  0.7340 & 0.1207 & (0.5710; 0.9610) &$\eta_4^{(2)}$& 1.3690 & 0.1773 & (1.1220; 1.6980) \\ 
		$\eta_5^{(1)}$ & 0.9680 & 0.1461 & (0.7650; 1.2390) &$\eta_5^{(2)}$& 0.9980 & 0.1390 & (0.8010;1.2550) \\ 
		$\eta_6^{(1)}$ & 0.8260 & 0.1149 & (0.6580; 1.0310) &$\eta_6^{(2)}$&  1.4150 & 0.2656 & (1.0520; 1.9120) \\ 
		$\eta_7^{(1)}$ & 1.2260 & 0.1464 & (1.0140; 1.4880) &$\eta_7^{(2)}$& 1.5060 & 0.1727 & (1.2560; 1.8240) \\ 
		$\eta_8^{(1)}$ &0.8510 & 0.1276 & (0.6760; 1.0940) 
		&$\eta_8^{(2)}$& 1.2390 & 0.2478 & (0.9200; 1.7170) \\ 
		$\eta_9^{(1)}$ & 0.7230 & 0.0772 & (0.6060; 0.8600 ) &$\eta_9^{(2)}$& 1.2450 & 0.2061 & (0.9440; 1.6230) \\ 
		$\eta_{10}^{(1)}$ & 0.2200 & 0.0592 & (0.1480; 0.3360) &$\eta_{10}^{(2)}$& 1.7260 & 0.3244 & (1.2560; 2.3350) \\ 
		\hline 
		$\nu_1^{(1)}$ & 0.0160 & 0.0022 & (0.0130; 0.0200) 
		& $\nu_1^{(2)}$&0.0590 & 0.0080 & (0.0470; 0.0740) \\ 
		$\nu_2^{(1)}$ & 0.1430 & 0.0192 & (0.1140; 0.1770) 
		& $\nu_2^{(2)}$& 0.1050 & 0.0139 & (0.0840; 0.1300) \\ 
		$\nu_3^{(1)}$ & 0.1270 & 0.0136 & (0.1060; 0.1510 ) 
		& $\nu_3^{(2)}$& 0.0670 & 0.0095 & (0.0530; 0.0840) \\ 
		$\nu_4^{(1)}$ & 0.0930 & 0.0111 & (0.0760; 0.1120)
		& $\nu_4^{(2)}$& 0.0310 & 0.0040 & (0.0250; 0.0390) \\ 
		$\nu_5^{(1)}$ & 0.1190 & 0.0153 & (0.0970; 0.1470) 
		& $\nu_5^{(2)}$&  0.0300 & 0.0032 & (0.0250; 0.0350) \\ 
		$\nu_6^{(1)}$ & 0.0510 & 0.0088 & (0.0380; 0.0670) 
		& $\nu_6^{(2)}$& 0.0160 & 0.0019 & (0.0130; 0.0190) \\ 
		$\nu_7^{(1)}$ & 0.0400 & 0.0065 & (0.0310; 0.0520) 
		& $\nu_7^{(2)}$& 0.0150 & 0.0018 & (0.0120; 0.0180 ) \\ 
		$\nu_8^{(1)}$ & 0.0100 & 0.0010 & (0.0090; 0.0120) 
		& $\nu_8^{(2)}$& 0.0050 & 0.0008 & (0.0040; 0.0070) \\ 
		$\nu_9^{(1)}$ & 0.0200 & 0.0040 & (0.0140; 0.0270) 
		& $\nu_9^{(2)}$&0.0020 & 0.0003 & (0.0020; 0.0030 ) \\ 
		$\nu_{10}^{(1)}$ & 0.0050 & 0.0008 & (0.0040; 0.0060) 
		& $\nu_{10}^{(2)}$& 0.0030 & 0.0004 & (0.0020; 0.0030 ) \\ 
		\hline
		$\gamma^{(1)}$ & 0.1400 & 0.0372 & (0.0900; 0.2120) & $\gamma^{(2)}$  & 0.1580 & 0.0431 & (0.1030; 0.2430) \\ 
		$\delta$ & 0.3240 & 0.0732 & (0.2220; 0.4610) &$p$ & 1.8290 & 0.0660 & (1.7120; 1.9260)  \\
		\hline
		\bottomrule
	\end{tabular}
	\caption{Posterior statistics of parameters from marginal estimation}\label{Tab:marreal}
	
\end{table}

\begin{table}[h!]
	\centering
	\begin{tabular}{lccc}
		\toprule
		\hline
		& Median & SD & 90\% CI \\ 
		\hline
		$c$ & 1.0080 & 4.6868 & (0.0570; 17.2280) \\ 
		$\beta$ & 3.0910 & 0.9413 & (1.8920; 5.0220) \\ 
		\hline
		\bottomrule
	\end{tabular}
	\caption{Posterior statistics of parameters from multivariate estimation}\label{Tab:mulreal}
\end{table}

\end{document}